\documentclass{article}

\usepackage{arxiv}

\usepackage[utf8]{inputenc} 
\usepackage[T1]{fontenc}    
\usepackage{hyperref}       
\usepackage{url}            
\usepackage{booktabs}       
\usepackage{amsfonts}       
\usepackage{nicefrac}       
\usepackage{microtype}      
\usepackage{graphicx}
\usepackage{natbib}
\usepackage{doi}

\usepackage{xcolor}
\hypersetup{
  colorlinks=true,
  linkcolor=red,
  citecolor=orange,
  urlcolor=blue,
}

\graphicspath{{./figure}}

\title{Large-eddy simulation of turbulent separated and reattached flow in enlarged annular pipe}

\author{\href{https://orcid.org/0000-0002-4875-8174}{\includegraphics[scale=0.06]{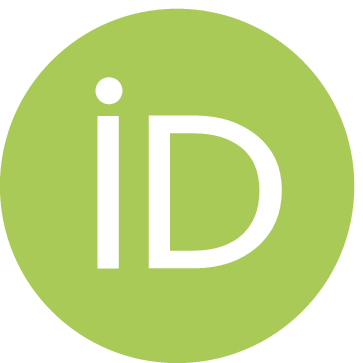}\hspace{1mm}Hideki Yanaoka}
\thanks{Email address for correspondence: yanaoka@iwate-u.ac.jp} \\
	Department of Systems Innovation Engineering, \\
    Faculty of Science and Engineering, Iwate University, \\
    4-3-5 Ueda, Morioka1, Iwate 202-8551, Japan \\
	\And
	Naoto Yamada \\
	Hokkaido Electric Power Company,Incorporated, \\
	2, Higashi 1-chome, Odori, Chuo-ku, Sapporo, Hokkaido 060-8677, Japan
	\texttt{} \\
}

\renewcommand{\shorttitle}{Turbulent separated and reattached flow in enlarged annular pipe}

\makeatletter
\hypersetup{
pdfusetitle,
bookmarksnumbered,
pdftitle={\@title},
pdfsubject={},
pdfauthor={Hideki Yanaoka, Naoto Yamada},
pdfkeywords={Separation, Reattachment, Turbulent flow, Vortex, Annular pipe, Numerical simulation},
}
\makeatother

\begin{document}

\maketitle

\begin{abstract}
This study performs a large-eddy simulation of turbulent separated 
and reattached flow in an enlarged annular pipe. 
A vortex ring is periodically shed from the sudden expansion part. 
A longitudinal vortex occurs around the vortex ring, 
making the flow three-dimensional. 
As a result, the vortex ring becomes unstable downstream 
and splits into small vortices. 
A tubular longitudinal vortex structure occurs downstream of the reattachment point 
near the wall surface on the inner pipe side. 
A low-frequency fluctuation occurs at each pipe diameter ratio. 
The smaller the pipe diameter ratio is, 
the more downstream the influences of small-scale vortices and low-frequency fluctuation 
on the flow field appear. 
The smaller the pipe diameter ratio, the slower the pressure recovery 
downstream from the reattachment point. 
The pressure recovery on the inner pipe side is delayed compared to 
the outer pipe side. 
Turbulence is maximum upstream of the reattachment point due to 
the small-scale vortices generated by the collapse of the vortex ring. 
This maximum value decreases as the pipe diameter ratio decreases. 
The smaller the pipe diameter ratio, the higher the turbulence 
downstream from the reattachment point.
\end{abstract}


\section{Introduction}

Separated and reattached flows occur in various fluid machines and 
heat exchangers, 
causing deterioration in performance and efficiency and vibration and noise. 
On the other hand, since the heat transfer performance is improved by 
the strong mixing action of the flow field, the separated flow is actively used. 
Many studies have been conducted to clarify the characteristics of 
such separated flows 
\citep{Lane&Loehrke_1980, Ota_et_al_1981, Kiya&Sasaki_1983, Ayukawa_et_al_1985, Bruno_et_al_2010, Sasaki&Kiya_1991, Yanaoka&Ota_1996(a), Yanaoka&Ota_1996(b), Yanaoka&Ota_1996(c), Hussein_et_al_2011, Oon_et_al_2013, Oon_et_al_2014, Cao_et_al_2014, Hussein_et_al_2016}.

The flow in the annular pipe targeted in this study is the flow form seen 
in engineering application equipment such as heat exchangers, 
nuclear reactors, and combustion engines. 
Therefore, research on annular flow has been conducted. 
\citet{Chung_et_al_2002} performed direct numerical simulations 
on turbulent flow in an annular channel 
and found that in fully developed turbulent flow, 
velocity and turbulence distributions near the wall differ 
between the inner and outer tube sides. 
\citet{Rouiss_et_al_2009} clarified the effect of the heat flux ratio 
on the heat transfer characteristics in turbulent flow in the annular channel 
by changing the heat flux applied to the inner and outer pipes.

In an annular flow, a separated and reattached flow occurs due to 
rapid expansion or contraction of the flow channel. 
\citet{Hussein_et_al_2011} experimentally investigated the heat transfer 
characteristics in an enlarged annular flow, 
and \citet{Oon_et_al_2013, Oon_et_al_2014} investigated them 
by numerical analysis. 
On the other hand, \citet{Hussein_et_al_2016} investigated 
the heat transfer characteristics in an enlarged annular flow channel 
using a nanofluid as a working fluid. 
In this way, several studies have been conducted focusing on the heat transfer 
characteristics of the enlarged annular flow. 
On the other hand, the unsteady characteristics of the flow field have not been 
investigated in detail. 
In particular, the effect of the diameter ratio of the inner and outer pipes 
on the separated and reattached flow has not been sufficiently investigated. 
In actual equipment, annular tubes with various pipe diameter ratios are used. 
In addition, the distributions of velocity and turbulence 
in the separated and reattached flow observed 
in the enlarged annular flow channel is considered 
to be greatly affected by the pipe diameter ratio, 
as in the case of the turbulent flow in the fully developed annular flow channel.

In this study, we perform numerical analysis using the LES model 
for the turbulent separated and reattached flow in an enlarged annular pipe 
and clarify the influence of the pipe diameter ratio 
on the vortex structure and turbulence characteristics.

\section{Numerical procedures}

\begin{figure}[!t]
\centering
\includegraphics[trim=10mm 5mm 10mm 7mm, clip, width=110mm]{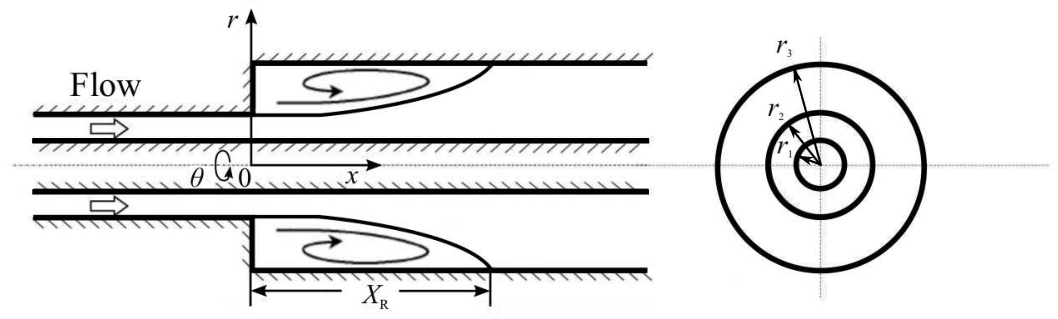}\\
\caption{Flow configuration and coordinate system}
\label{flow_model}
\end{figure}

Figure \ref{flow_model} shows the flow configuration and coordinate system. 
The origin is on the center axis at the sudden expansion step, 
and the $x$-, $y$-, and $z$-axes are the streamwise, radial, 
and circumferential directions, respectively. 
The velocities in these directions are denoted as $u_x$, $u_r$, and $u_\theta$, respectively. 
The inner radius of the annular pipe is $r_1$, 
the outer radius upstream of the step is $r_2$, 
and the outer radius downstream of the step is $r_3$. 
The hydraulic diameter $D$ is defined as $D=4A/L$. 
Here, $A$ is the cross-section area of the flow channel at the inlet, 
and $L$ is defined as $L=2 \pi (r_1 + r_2)$. 
In this study, we consider a turbulent flow field in an enlarged annular pipe 
with an expansion rate of $ER=2.0$, 
which is defined as $ER=(r_3-r_1)/(r_2-r_1)$.

This study deals with the three-dimensional flow of an incompressible viscous fluid 
with constant physical properties. 
The governing equations are the continuity and Navier-Stokes equations 
in a cylindrical coordinate system. 
These equations are non-dimensionalized using the hydraulic diameter $D$ 
and average velocity $U_\mathrm{in}$ at the inlet. 
The dimensionless governing equation is given as
\begin{equation}
   \nabla \cdot {\bf u} = 0,
\end{equation}
\begin{equation}
   \frac{\partial {\bf u}}{\partial t} + \nabla \cdot ({\bf u} \otimes {\bf u}) 
   = - \nabla p + \frac{1}{Re} \nabla \cdot (2 {\bf S}),
\end{equation}
where the strain rate tensor is defined as
\begin{equation}
   {\bf S} = \frac{1}{2} [ \nabla {\bf u} + (\nabla {\bf u})^T ],
\end{equation}
where coordinates and velocity vector are expressed as ${\bf x}=(x, r, \theta)$ 
and ${\bf u}=(u_x, u_r, u_\theta)$, respectively. 
$t$ is time, $p$ is the pressure, $Re=U_\mathrm{in} D/\nu$ is the Reynolds number, 
and $\nu$ is the kinematic viscosity.

This study uses the dynamic SGS model \citep{Germano_et_al_1991, Lilly_1992} 
as the LES model to analyze the turbulent field at a high Reynolds number. 
As a test filter, a top hat filter based on the trapezoidal rule is used 
\citep{Najjar&Tafti_1996}, 
and the ratio of the test filter width to the grid filter width is $\sqrt{6}$.

The governing equations are solved using the simplified marker and cell method 
\citep{Amsden&Harlow_1970}. 
The Crank-Nicolson method is applied to discretize the time derivative 
and then time marching is performed. 
The second-order central difference scheme is used to discretize 
the spatial derivative.

\section{Calculation conditions}

In this study, the inner radius of the annular pipe is set to be 
$r_1/D=0.15$, 0.3, and 0.5, 
and the outer radius upstream of the step is set to be 
$r_2/D=0.65$, 0.8, and 1.0. 
At that time, the pipe diameter ratios defined as 
$\alpha=r_1/r_2$ are $\alpha=0.23$, 0.375, and 0.5. 
We investigate the effect of changes in the diameter ratio on the flow field. 
The calculation area is from $x=-10D$ to $x=25D$ in the $x$-direction.

In turbulent flow analysis, it is a significant problem to give 
a turbulent inflow velocity at the inlet of a flow channel. 
This study generates a periodic flow in the calculation region 
from $x/D=-10$ to $x/D=-5.0$. 
At each time step, we use the velocity of the cross-section at $x/D=-5.0$ 
as the turbulent inflow velocity at the inlet of the annular pipe. 
At that time, we adjust the inflow velocity $u_x$ so that the flow rate 
at the inlet becomes constant. 
In addition, non-slip conditions are given on the wall surface, 
and convection boundary conditions are used at the outlet.

\begin{figure}[!t]
\centering
\includegraphics[trim=0mm 0mm 0mm 0mm, clip,width=80mm]{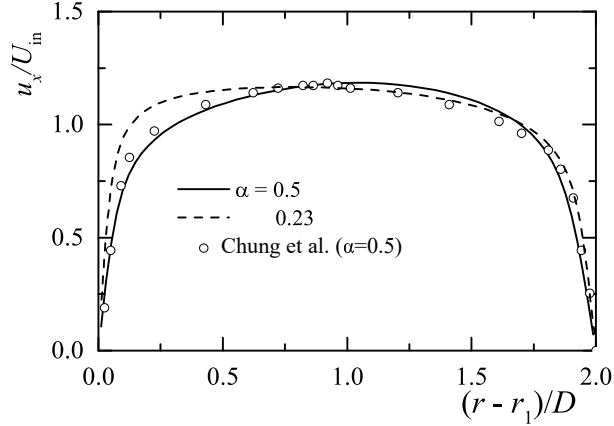} \\
\caption{Time--averaged velocity distribution at inlet boundary}
\label{inlet}
\end{figure}

The time-averaged streamwise velocity distribution at the inlet 
is compared with the previously reported result \citep{Chung_et_al_2002} 
in Fig. \ref{inlet} to validate the turbulent inflow velocity 
given at the inlet. 
Since this calculation result for $\alpha=0.5$ agrees well with the previous result, 
it can be seen that the turbulent flow is appropriately given in this calculation. 
Also, as in the previous study, when $\alpha$ becomes small, 
the peak of the distribution tends to be close to the inner pipe side 
of the annular pipe.

Three grids are used to confirm the dependency of the grids 
on the calculation results. 
The grid points of grid1, grid2, and grid3 are $254\times50\times58$, 
$311\times84\times61$, and $402\times114\times71$, respectively. 
The grid width is dense near the wall surface and the sudden expansion part. 
The minimum grid widths in grid1, grid2, and grid3 are $0.02D$, $0.01D$, 
and $0.005D$, respectively. 
We investigated the difference in calculation results depending on 
the number of grid points for $\alpha=0.5$, 
where the grid resolution in the circumferential direction is 
the coarsest among all the conditions. 
We will discuss the grid dependency later. 
In this study, to clarify the turbulence due to the vortex structure 
in more detail, the results using grid3 are mainly shown.

This study performs the calculation under the Reynolds number $Re=10^4$ 
defined by $D$ and $U_\mathrm{in}$. 
The time intervals used in the calculation are 
$\Delta t/(D/U_\mathrm{in})=0.01$, 0.005, and 0.0025 
for grid1, grid2, and grid3, respectively. 
The dimensionless time to start sampling the data is $T=0$.

\section{Results and discussion}

\subsection{Vortex structure}

\begin{figure}[!t]
\begin{minipage}{0.49\linewidth}
\begin{center}
\includegraphics[trim=5mm 0mm 0mm 10mm, clip, width=80mm]{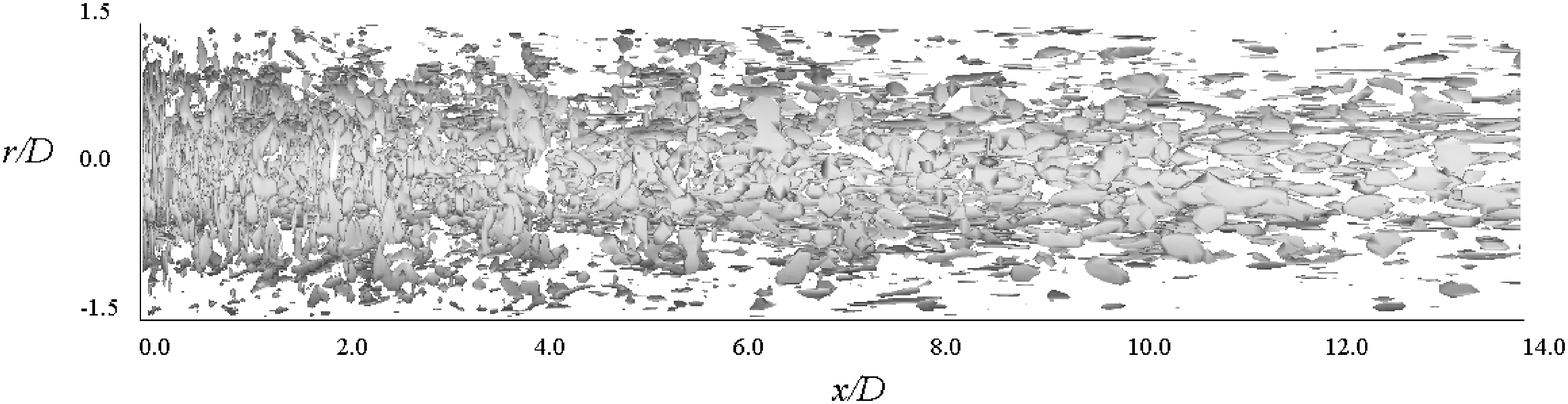} \\
(a) Side view
\end{center}
\end{minipage}
\begin{minipage}{0.49\linewidth}
\begin{center}
\includegraphics[trim=10mm 0mm 0mm 15mm, clip, width=65mm]{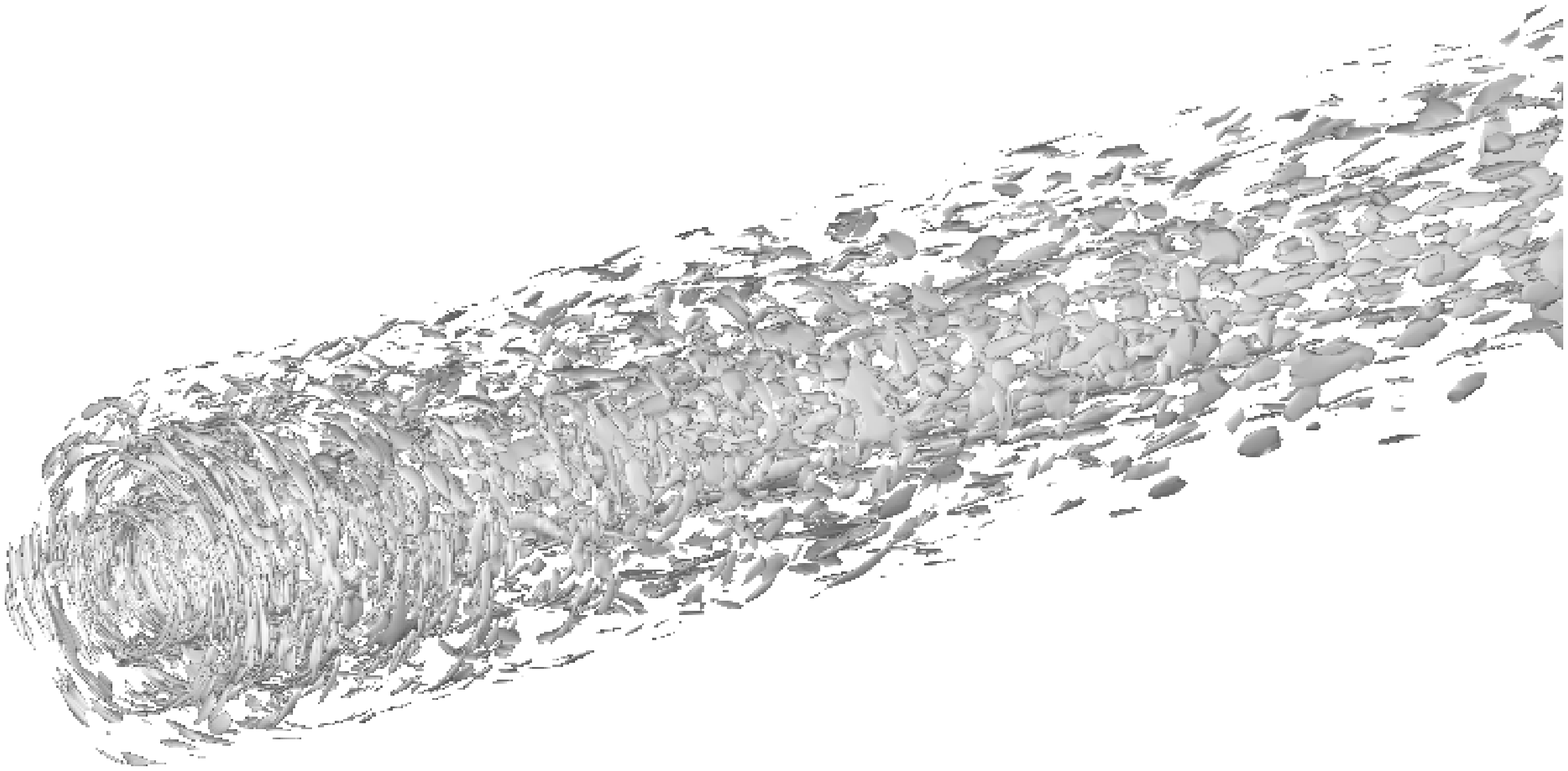} \\
\vspace*{-2.0\baselineskip}
(b) Perspective view
\end{center}
\end{minipage}
\caption{Isosurface of curvature of equipressure surface 
at $T=0$ for $\alpha=0.5$: Isosurface value is $-27/D$.}
\label{cur_r05}
\end{figure}

\begin{figure}[!t]
\begin{minipage}{0.49\linewidth}
\begin{center}
\includegraphics[trim=5mm 5mm 0mm 15mm, clip, width=80mm]{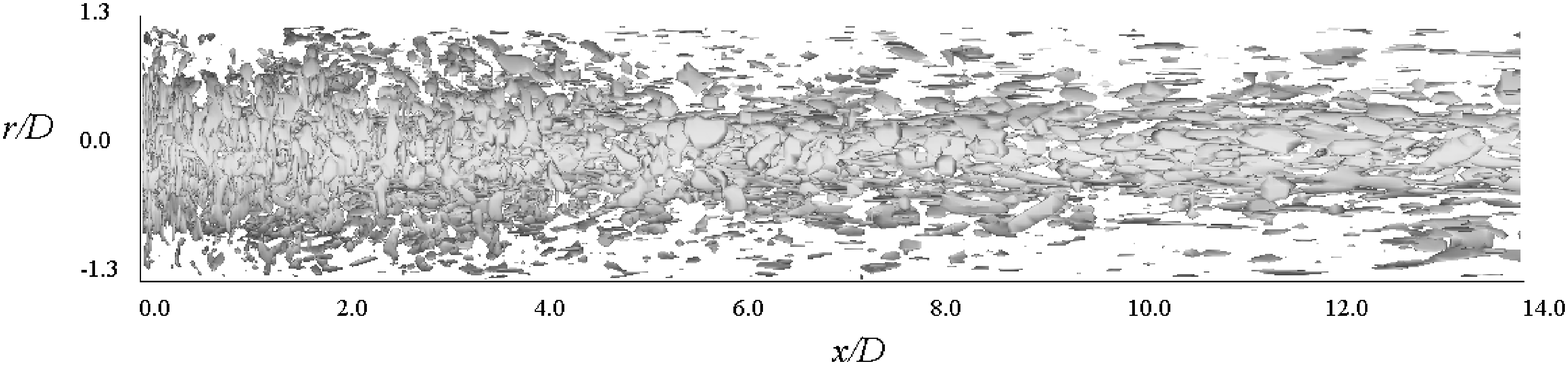} \\
(a) Side view
\end{center}
\end{minipage}
\begin{minipage}{0.49\linewidth}
\begin{center}
\includegraphics[trim=5mm 5mm 0mm 20mm, clip, width=65mm]{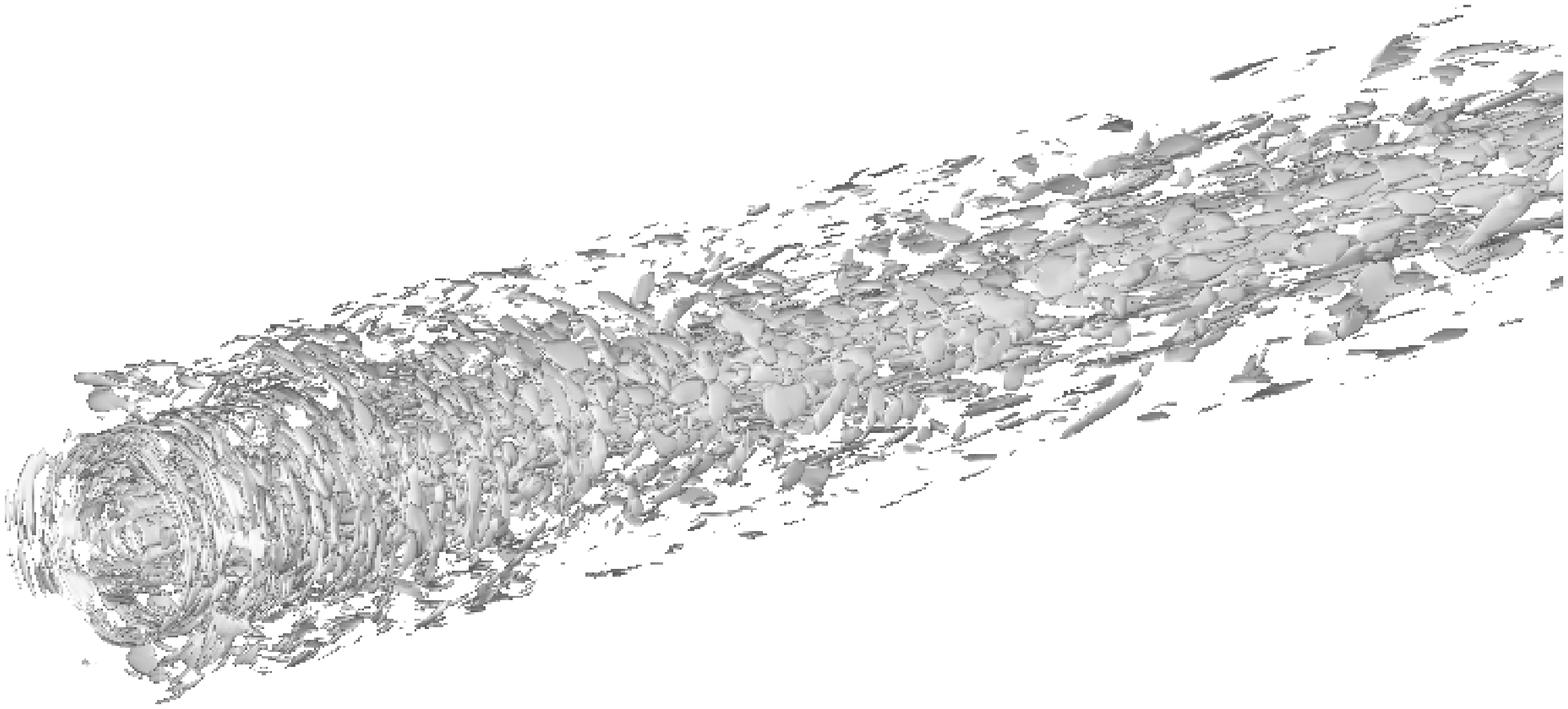} \\
\vspace*{-2.0\baselineskip}
(b) Perspective view
\end{center}
\end{minipage}
\caption{Isosurface of curvature of equipressure surface 
at $T=0$ for $\alpha=0.375$: Isosurface value is $-27/D$.}
\label{cur_r03}
\end{figure}

\begin{figure}[!t]
\begin{minipage}{0.49\linewidth}
\begin{center}
\includegraphics[trim=5mm 5mm 0mm 15mm, clip, width=80mm]{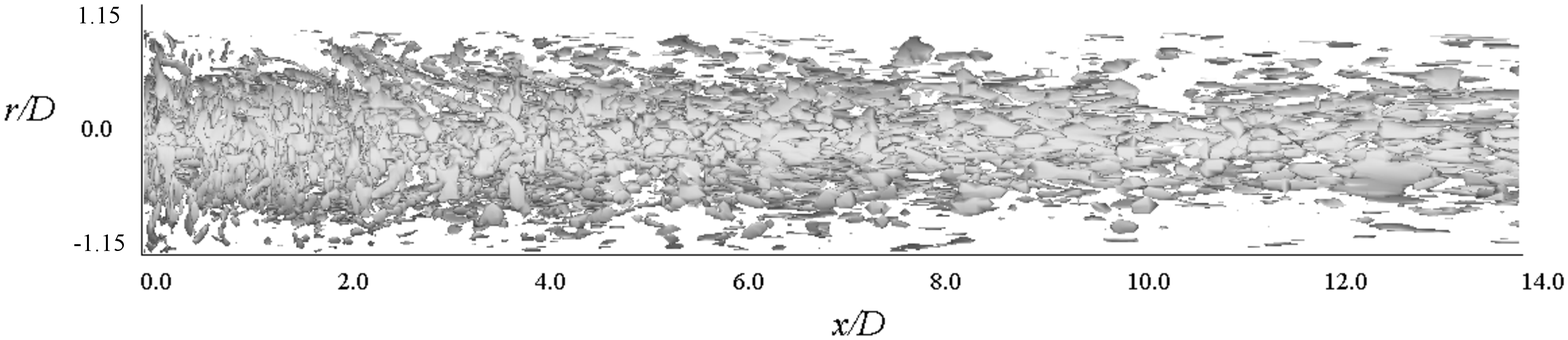} \\
(a) Side view
\end{center}
\end{minipage}
\begin{minipage}{0.49\linewidth}
\begin{center}
\includegraphics[trim=5mm 5mm 0mm 15mm, clip, width=65mm]{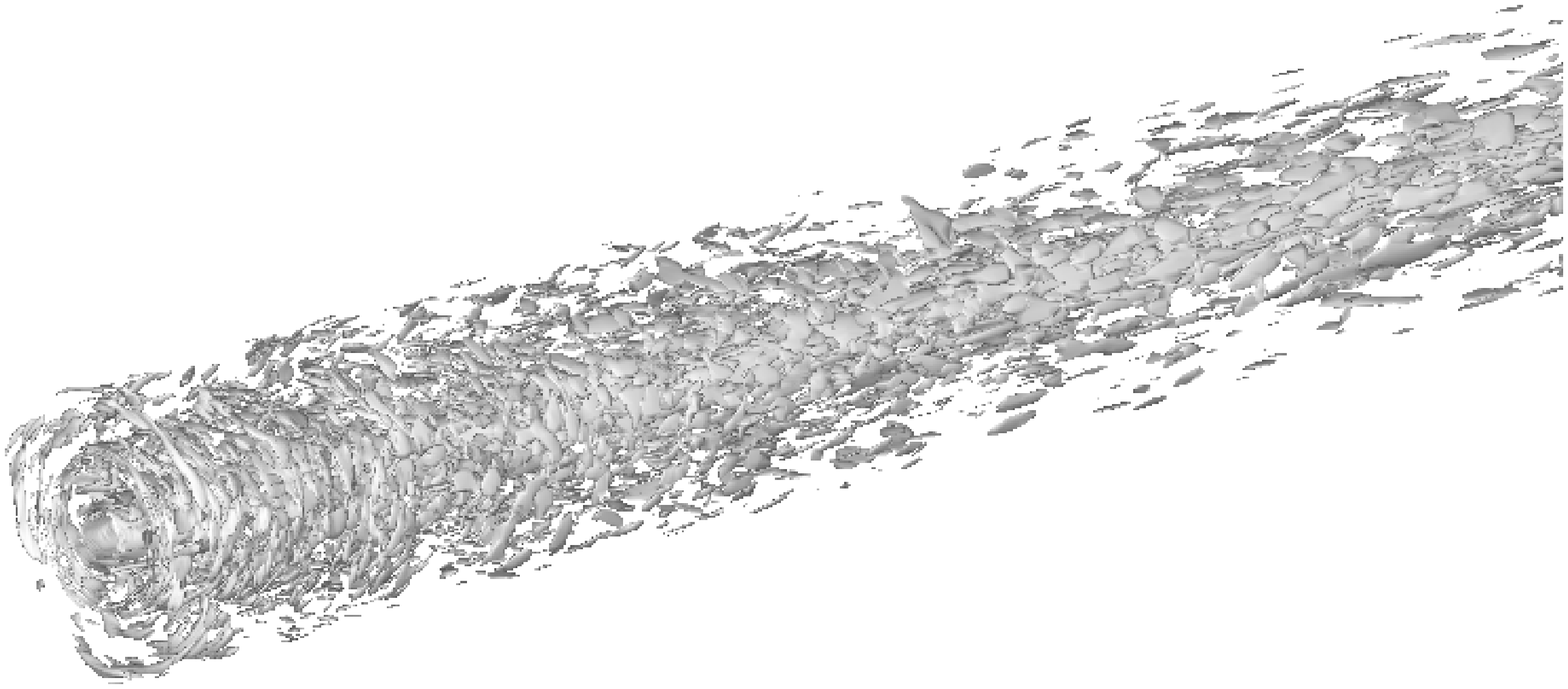} \\
\vspace*{-2.0\baselineskip}
(b) Perspective view
\end{center}
\end{minipage}
\caption{Isosurface of curvature of equipressure surface 
at $T=0$ for $\alpha=0.23$: Isosurface value is $-27/D$.}
\label{cur_r015}
\end{figure}

To clarify the vortex structure existing in the flow field, 
Fig. \ref{cur_r05} to Fig. \ref{cur_r015} show the isosurface of the curvature 
calculated from the equipressure surface. 
The reattachment point $X_\mathrm{R}$ is $X_\mathrm{R}=4.07D$ for $\alpha=0.5$, 
$X_\mathrm{R}=4.84$ for $\alpha=0.375$, and $X_\mathrm{R}=5.84D$ for $\alpha=0.23$. 
In all $\alpha$ values, the shear layer separated from the step of 
the sudden expansion step rolls up, forming a vortex ring periodically. 
This vortex ring has a deformed shape due to turbulence. 
The vortex ring becomes unstable downstream, 
and the vortex ring splits into small vortices around $x/D=3.0-4.0$ 
for $\alpha=0.5$, $x/D=3.5-4.5$ for $\alpha=0.375$, 
and $x/D=4.0-5.0$ for $\alpha=0.23$. 
Downstream from the reattachment point, 
a tubular vortex extending in the streamwise direction can be confirmed 
mainly over the region from the center of the flow channel 
to the wall surface on the inner pipe side. 
We can find from the above result that the smaller $\alpha$, 
the more downstream the deformation process of a series of vortices occurs.

%
%

\begin{figure}[!t]
\begin{minipage}{0.6\linewidth}
\begin{center}
\includegraphics[trim=1mm 0mm 8mm 10mm, clip, width=75mm]{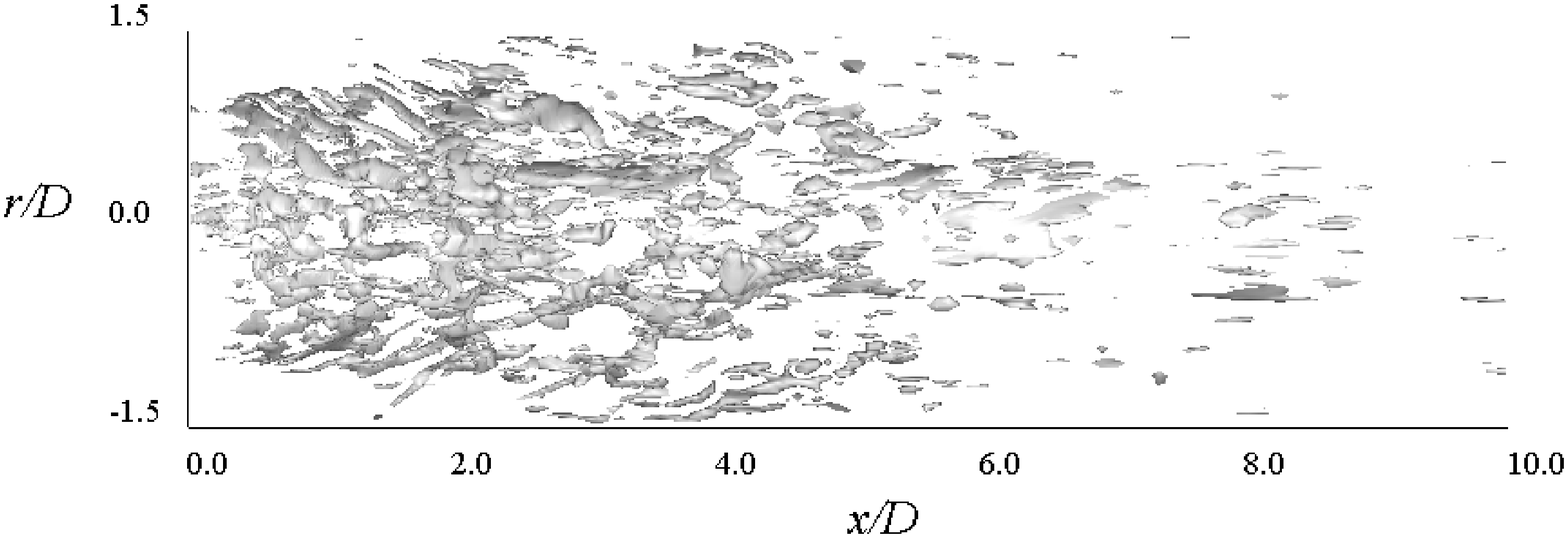} \\
(a) Isosurface: Isosurface value is $\omega_x=5.0U_\mathrm{in}/D$. \\
\includegraphics[trim=1mm 0mm 8mm 0mm, clip, width=75mm]{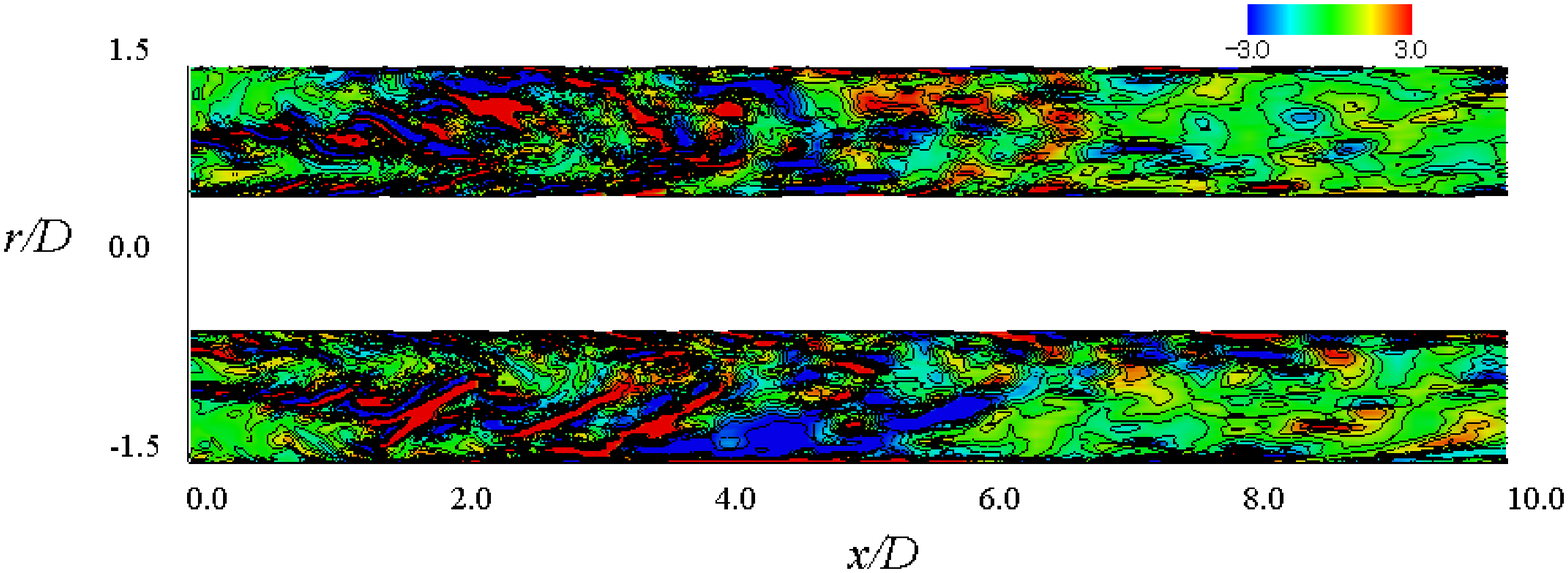} \\
(b) Contour in $x$-$r$ plane: Contour interval is 0.5 from $-5.0$ to 5.0.
\end{center}
\end{minipage}
\begin{minipage}{0.39\linewidth}
\begin{center}
\includegraphics[trim=5mm 0mm 0mm 0mm, clip, width=50mm]{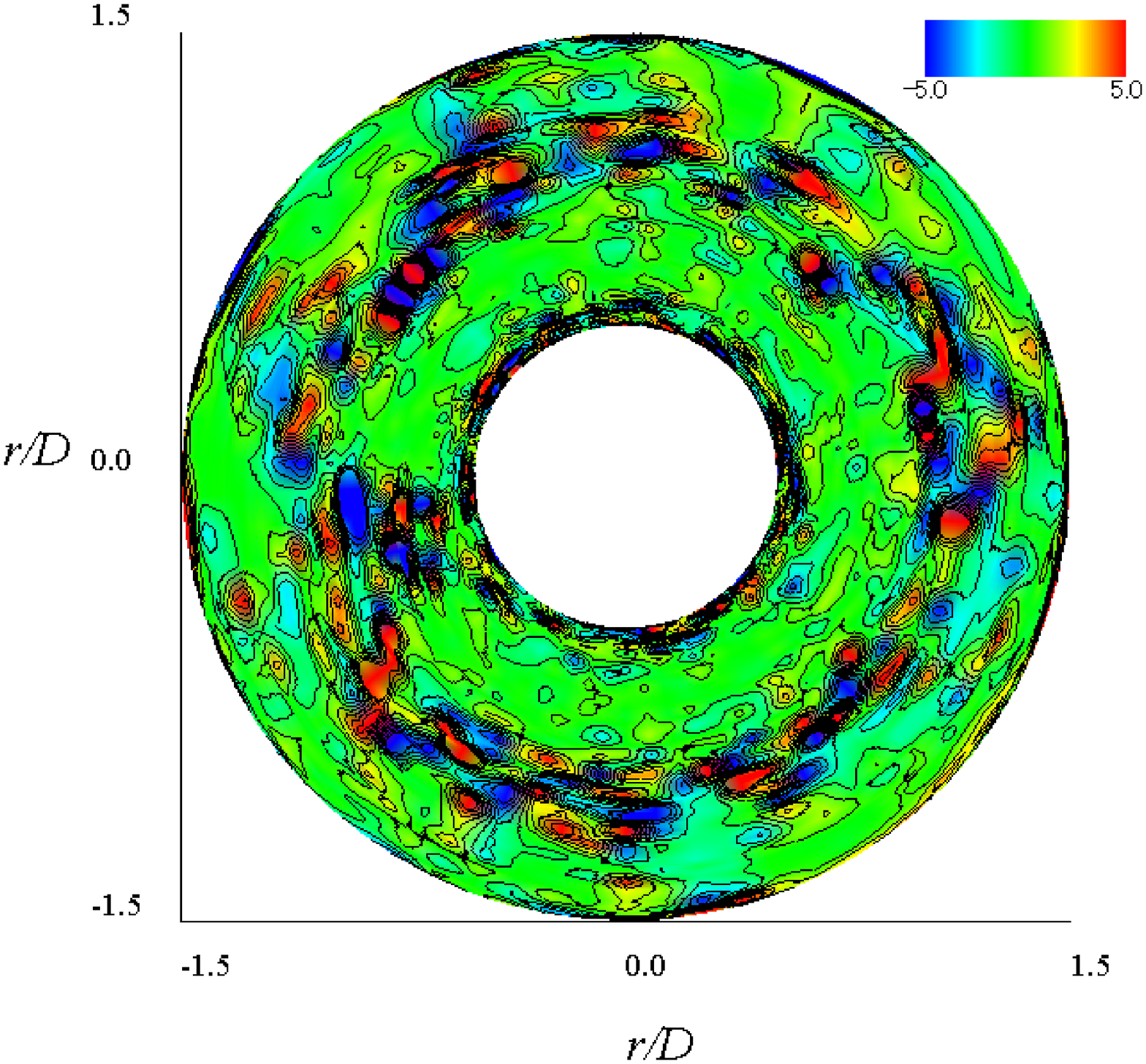} \\
(c) Contour in $r$-$\theta$ plane at $x/D=1.5$: Contour interval is 0.5 from $-5.0$ to 5.0.
\end{center}
\end{minipage}
\caption{Streamwise vorticity for $\alpha=0.5$.}
\label{vor_r05}
%
%
\begin{minipage}{0.6\linewidth}
\begin{center}
\includegraphics[trim=1mm 0mm 8mm 10mm, clip, width=75mm]{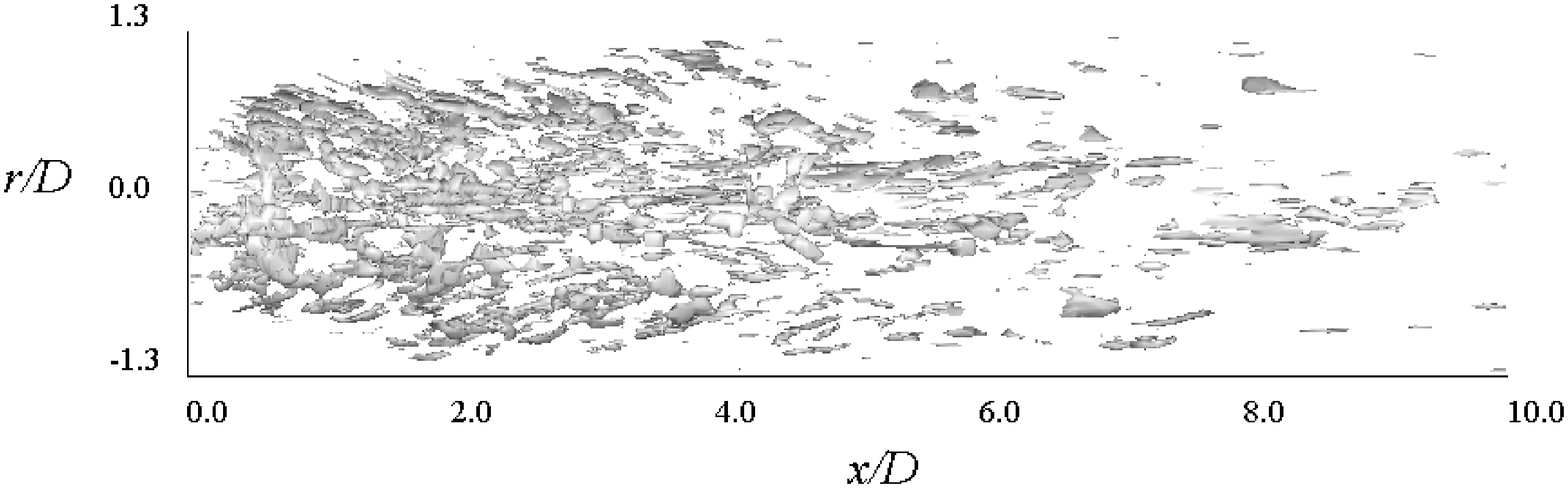} \\
(a) Isosurface: Isosurface value is $\omega_x=5.0U_\mathrm{in}/D$. \\
\includegraphics[trim=1mm 0mm 8mm 0mm, clip, width=75mm]{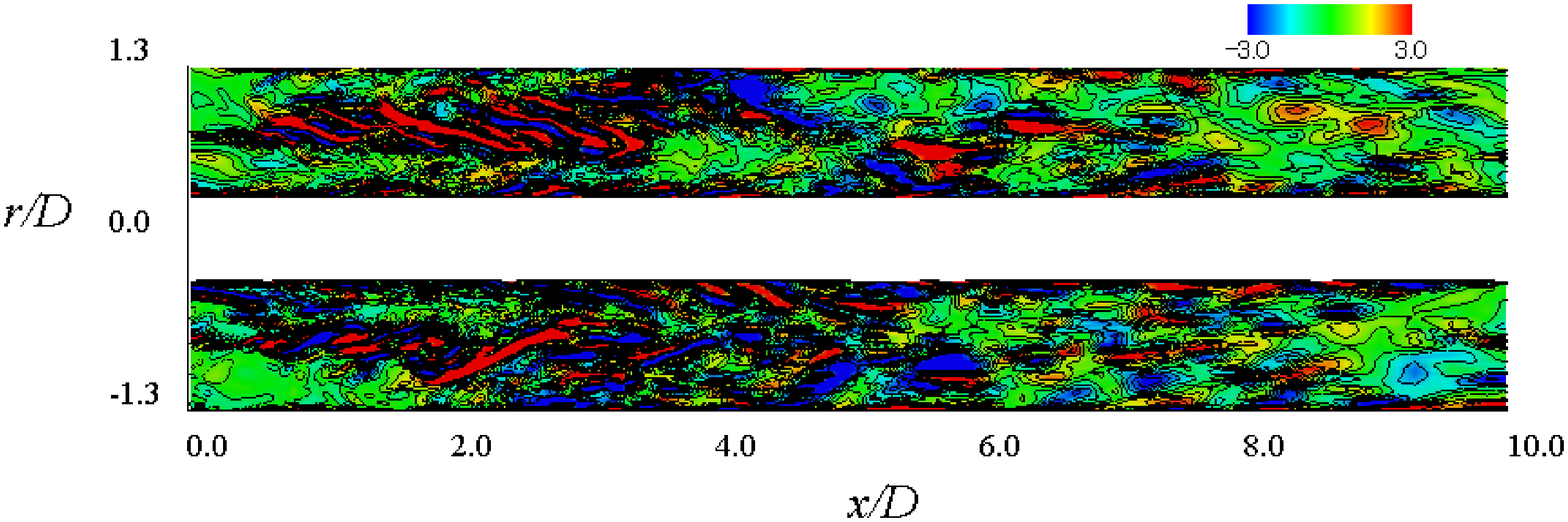} \\
(b) Contour in $x$-$r$ plane: Contour interval is 0.5 from $-5.0$ to 5.0.
\end{center}
\end{minipage}
\begin{minipage}{0.39\linewidth}
\begin{center}
\includegraphics[trim=5mm 0mm 0mm 0mm, clip, width=50mm]{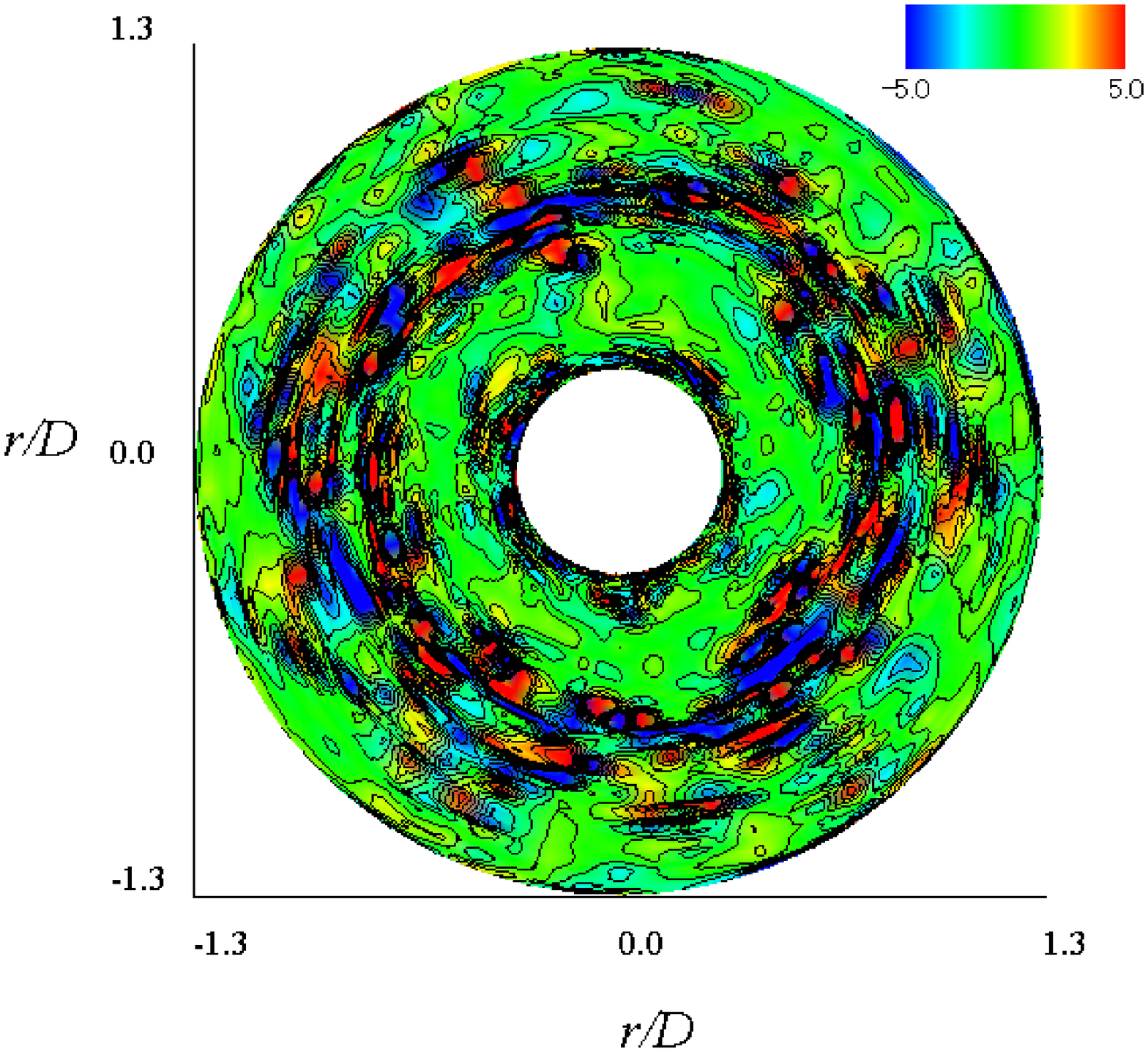} \\
(c) Contour in $r$-$\theta$ plane at $x/D=1.5$: Contour interval is 0.5 from $-5.0$ to 5.0.
\end{center}
\end{minipage}
\caption{Streamwise vorticity for $\alpha=0.375$.}
\label{vor_r03}
%
%
\begin{minipage}{0.6\linewidth}
\begin{center}
\includegraphics[trim=1mm 0mm 8mm 10mm, clip, width=75mm]{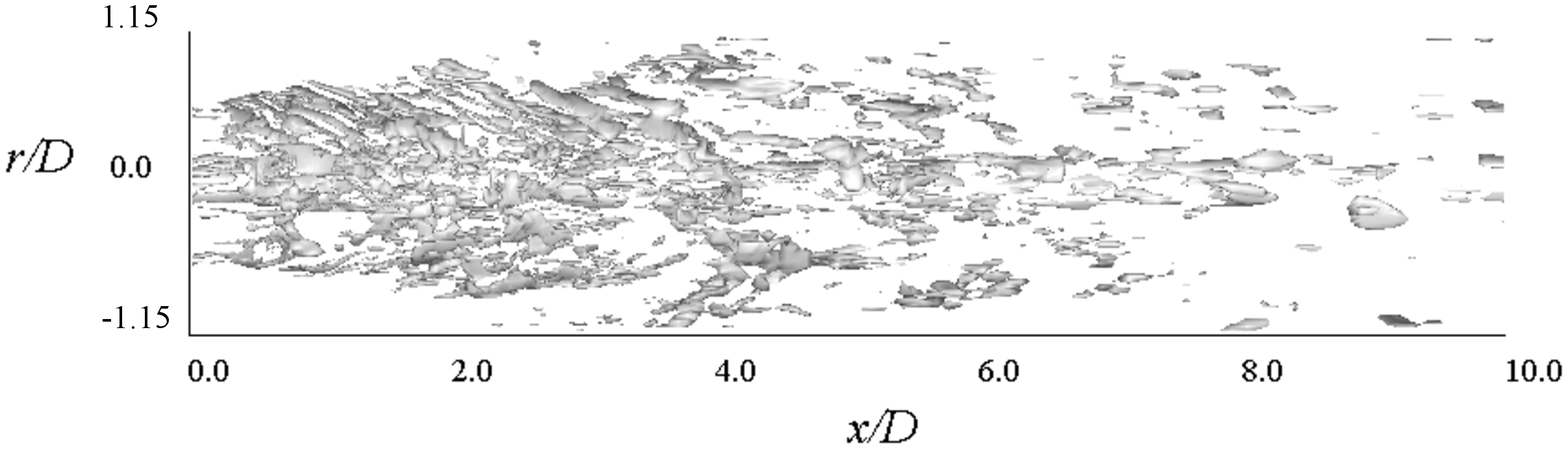} \\
(a) Isosurface: Isosurface value is $\omega_x=5.0U_\mathrm{in}/D$. \\
\includegraphics[trim=1mm 0mm 8mm 0mm, clip, width=75mm]{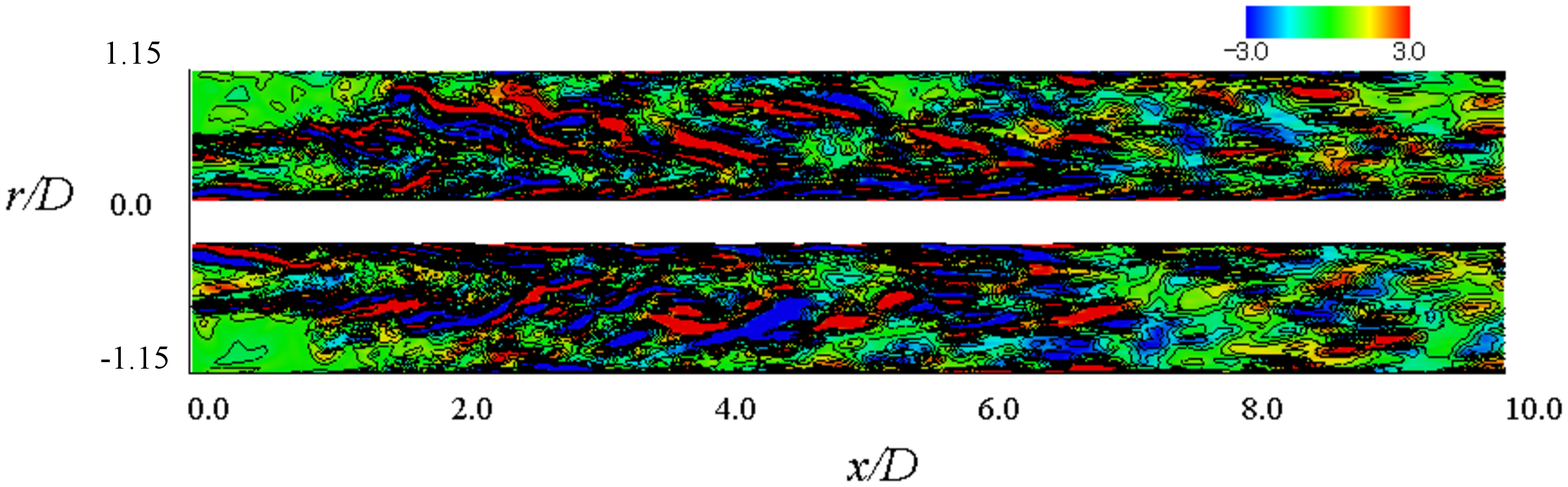} \\
(b) Contour in $x$-$r$ plane: Contour interval is 0.5 from $-5.0$ to 5.0.
\end{center}
\end{minipage}
\begin{minipage}{0.39\linewidth}
\begin{center}
\includegraphics[trim=5mm 0mm 0mm 0mm, clip, width=50mm]{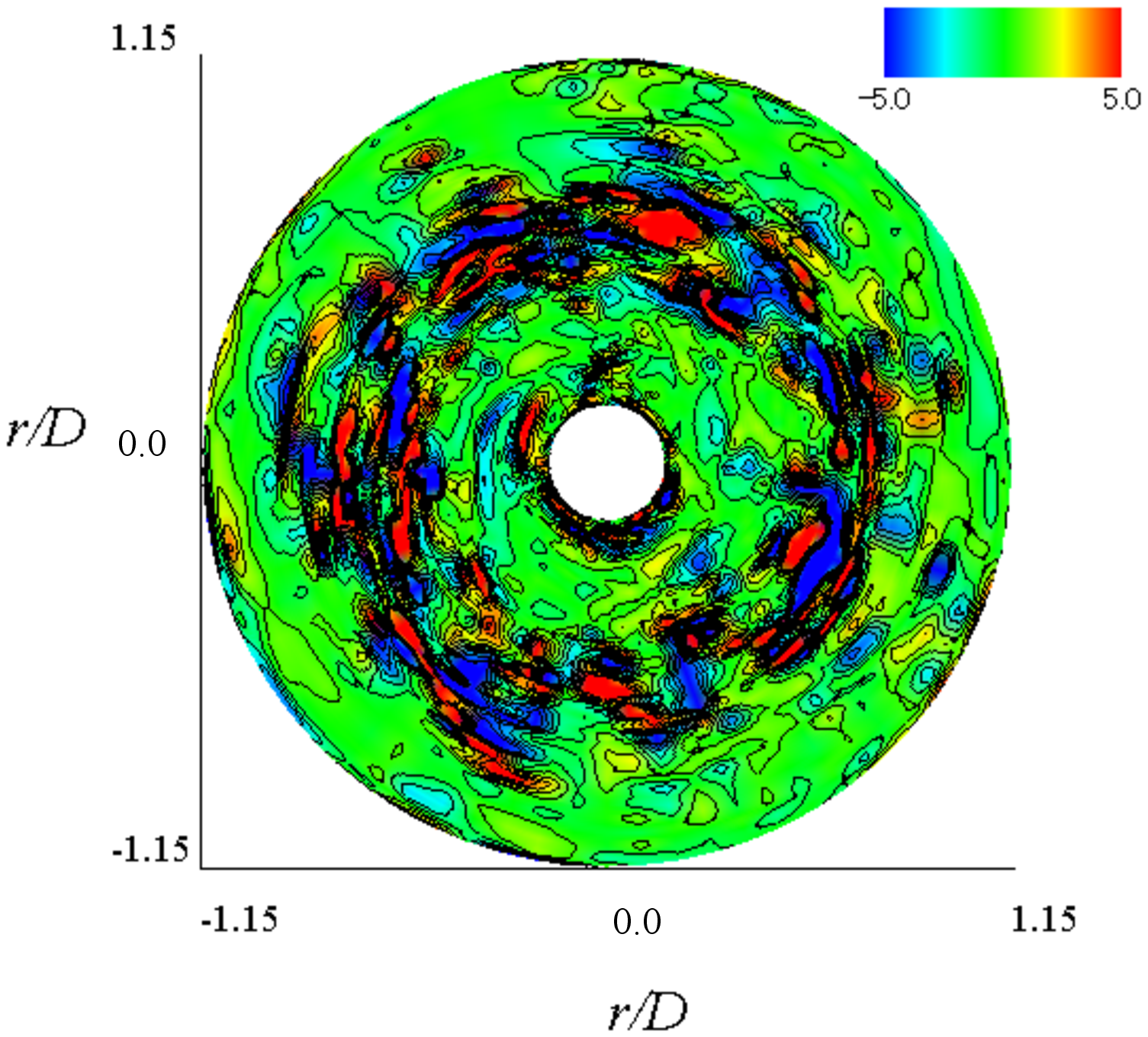} \\
(c) Contour in $r$-$\theta$ plane at $x/D=1.5$: Contour interval is 0.5 from $-5.0$ to 5.0.
\end{center}
\end{minipage}
\caption{Streamwise vorticity for $\alpha=0.23$.}
\label{vor_r015}
\end{figure}

Next, we clarify the longitudinal vortex structure for each $\alpha$ value in detail. 
Figures \ref{vor_r05} to \ref{vor_r015} show the isosurface of 
the streamwise vorticity, the contour in the $x$-$r$ plane on the central axis, 
and the contour in the $r$-$\theta$ plane at $x/D=1.5$. 
Regardless of $\alpha$, many longitudinal vortex structures exist 
around the vortex ring generated by the rollup of the shear layer. 
This vortex is similar to the rib structure connecting strong lateral vortices 
lined up in a turbulent mixing layer. 
This longitudinal vortex strengthens the mixing of the flow 
and makes the flow three-dimensional. 
Therefore, the vortex ring becomes unstable, collapses, 
and splits into small-scale vortices downstream. 
In addition, we can confirm that the smaller $\alpha$ is, 
the stronger the longitudinal vortex structure exists downstream.

\begin{figure}[!t]
\begin{minipage}{0.49\linewidth}
\begin{center}
\includegraphics[trim=3mm 0mm 0mm 10mm, clip, width=80mm]{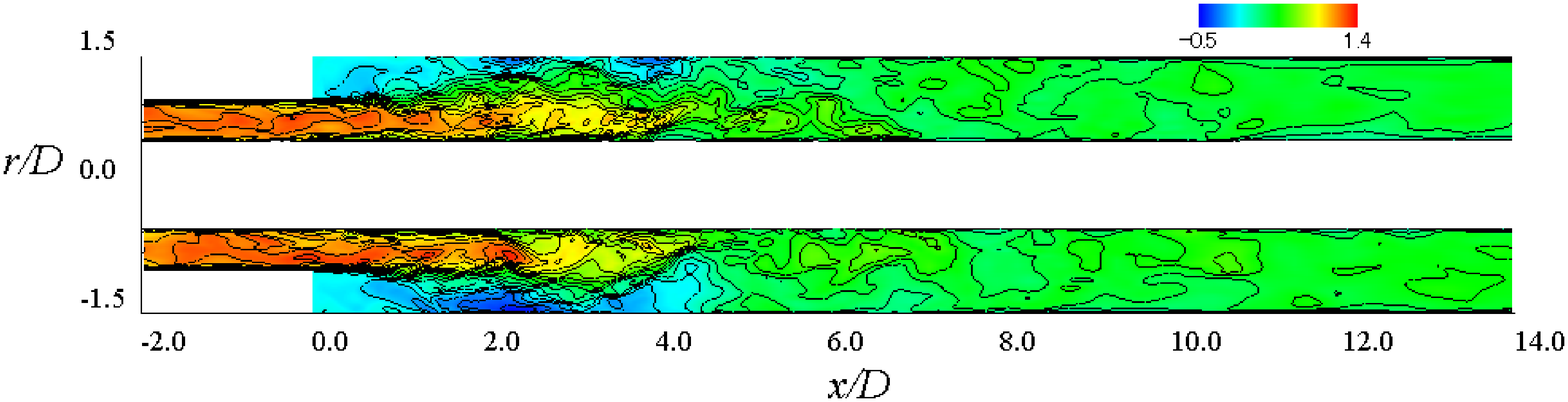} \\
(a) Streamwise velocity contours: Contour interval is 0.1 from $-0.5$ to 1.4. \\
\includegraphics[trim=0mm 0mm 0mm 10mm, clip, width=80mm]{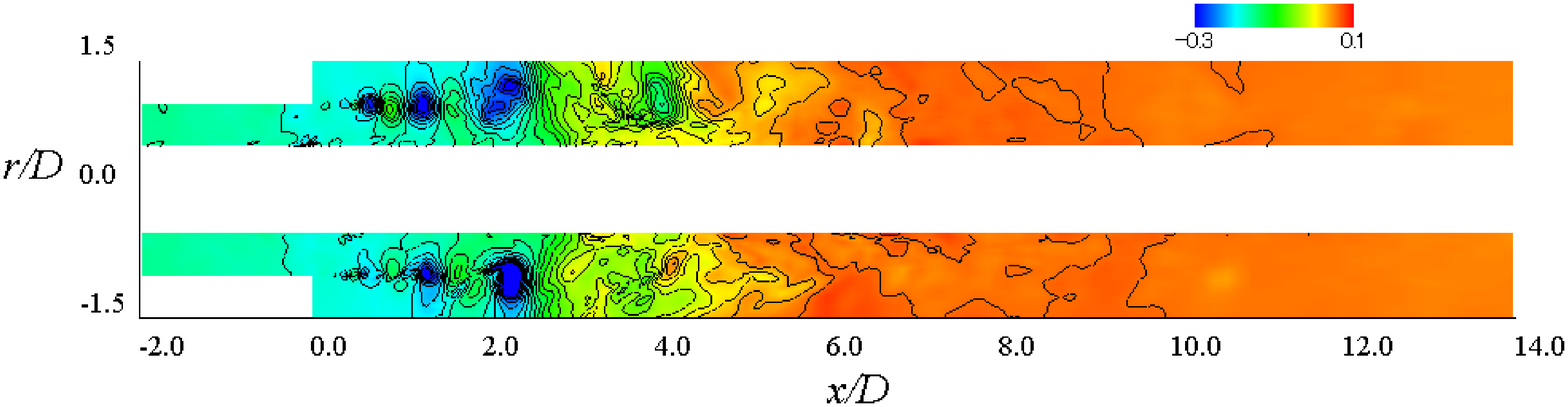} \\
(b) Pressure contours: Contour interval is 0.02 from $-3.0$ to 0.1. \\
\includegraphics[trim=0mm 0mm 0mm 10mm, clip, width=80mm]{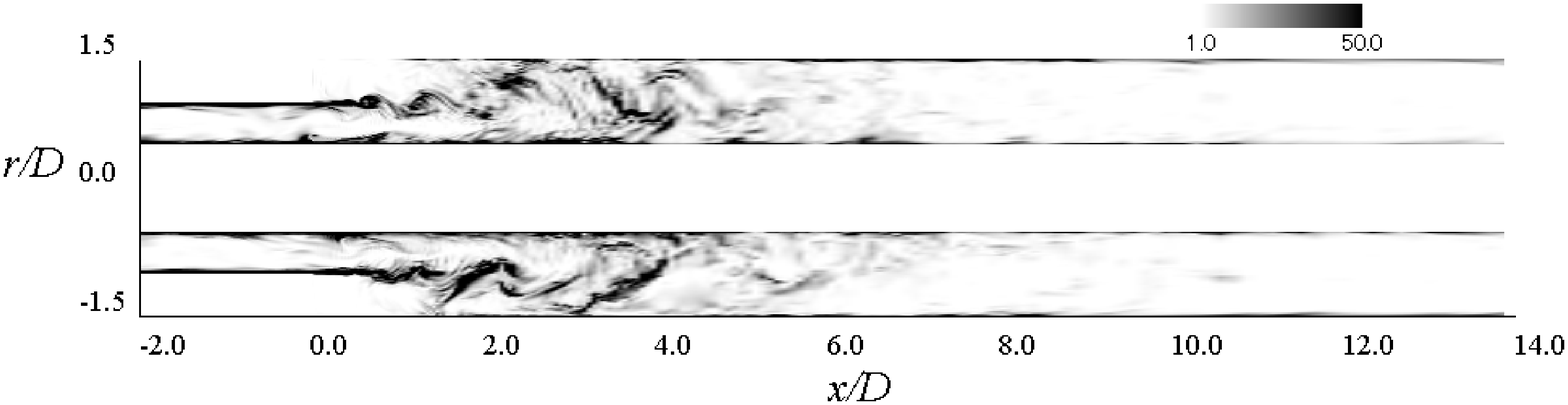} \\
(c) Enstrophy contours
\caption{Instantaneous flow field in $x$-$r$ plane for $\alpha=0.5$.}
\label{inst_r05}
\end{center}
\end{minipage}
%
%
\hspace*{0.02\linewidth}
\begin{minipage}{0.49\linewidth}
\begin{center}
\includegraphics[trim=3mm 0mm 0mm 10mm, clip, width=80mm]{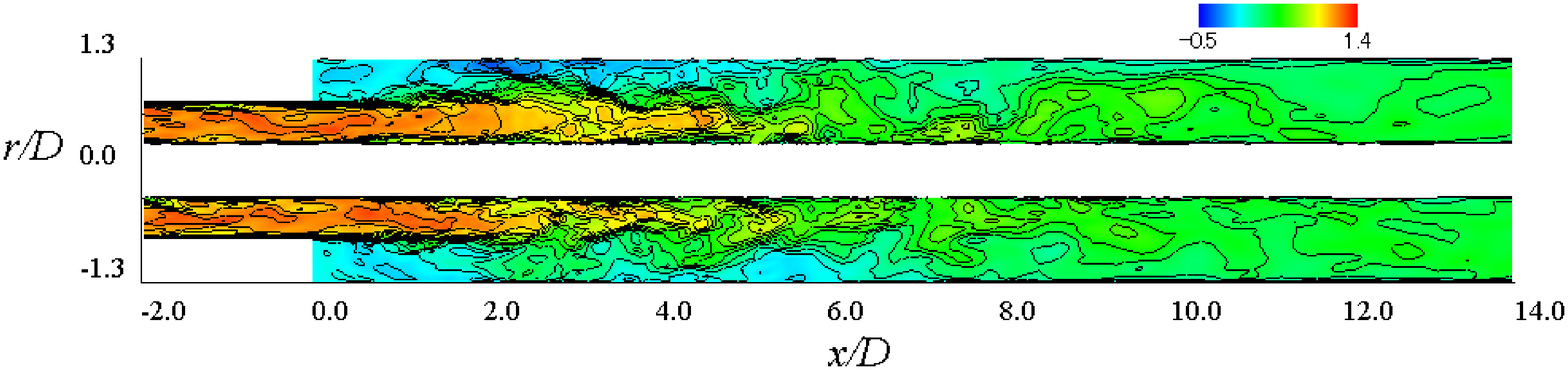} \\
(a) Streamwise velocity contours: Contour interval is 0.1 from $-0.5$ to 1.4. \\
\includegraphics[trim=0mm 0mm 0mm 10mm, clip, width=80mm]{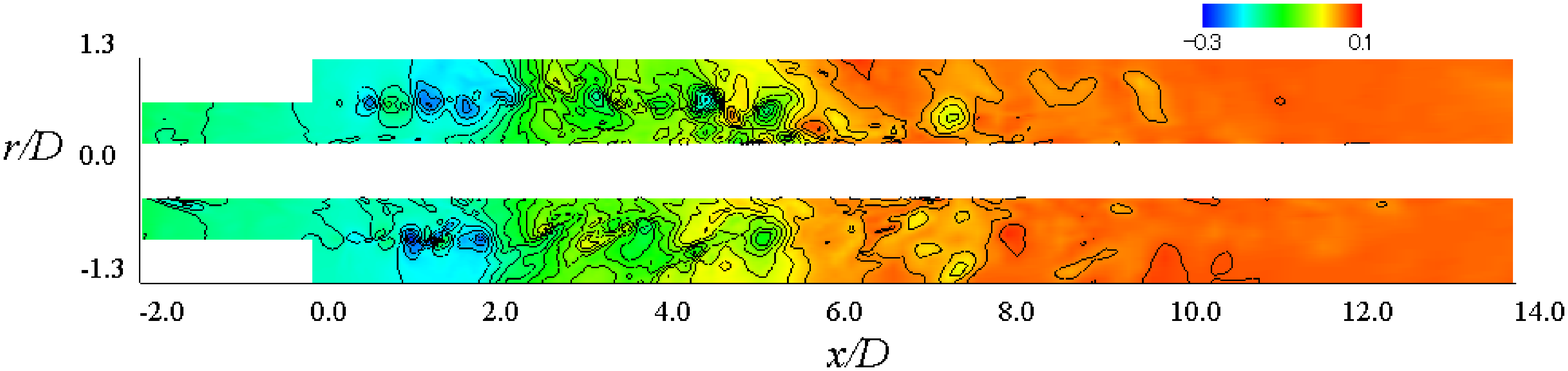} \\
(b) Pressure contours: Contour interval is 0.02 from $-3.0$ to 0.1. \\
\includegraphics[trim=0mm 0mm 0mm 10mm, clip, width=80mm]{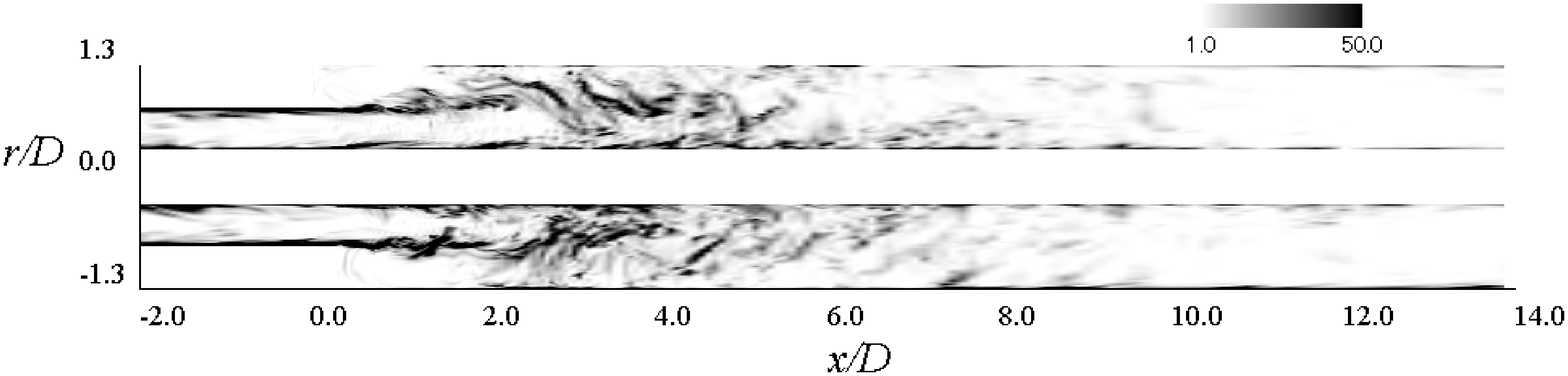} \\
(c) Enstrophy contours
\caption{Instantaneous flow field in $x$-$r$ plane for $\alpha=0.375$.}
\label{inst_r03}
\end{center}
\end{minipage}
%
%
%
\begin{center}
\includegraphics[trim=3mm 5mm 0mm 10mm, clip, width=80mm]{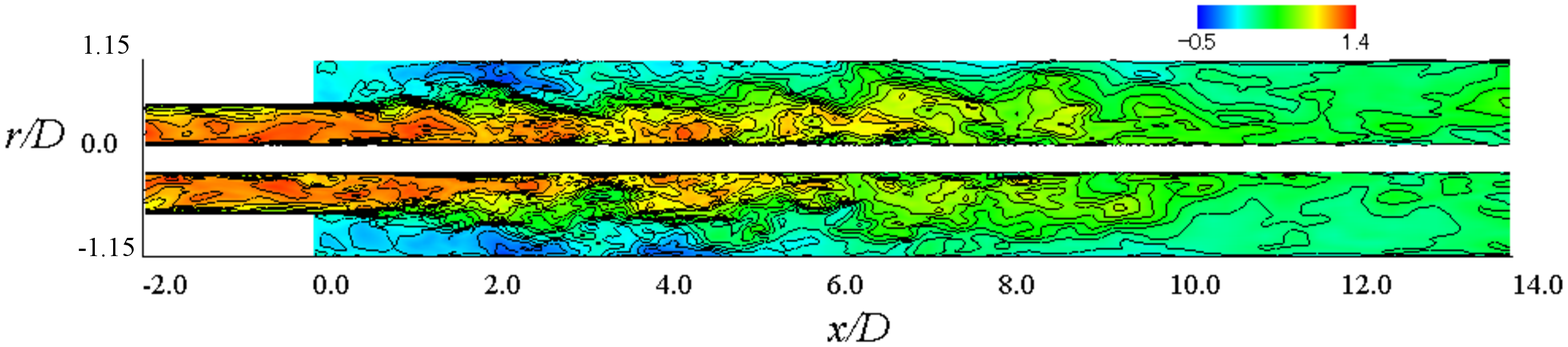} \\
(a) Streamwise velocity contours: Contour interval is 0.1 from $-0.5$ to 1.4. \\
\includegraphics[trim=0mm 5mm 0mm 10mm, clip, width=80mm]{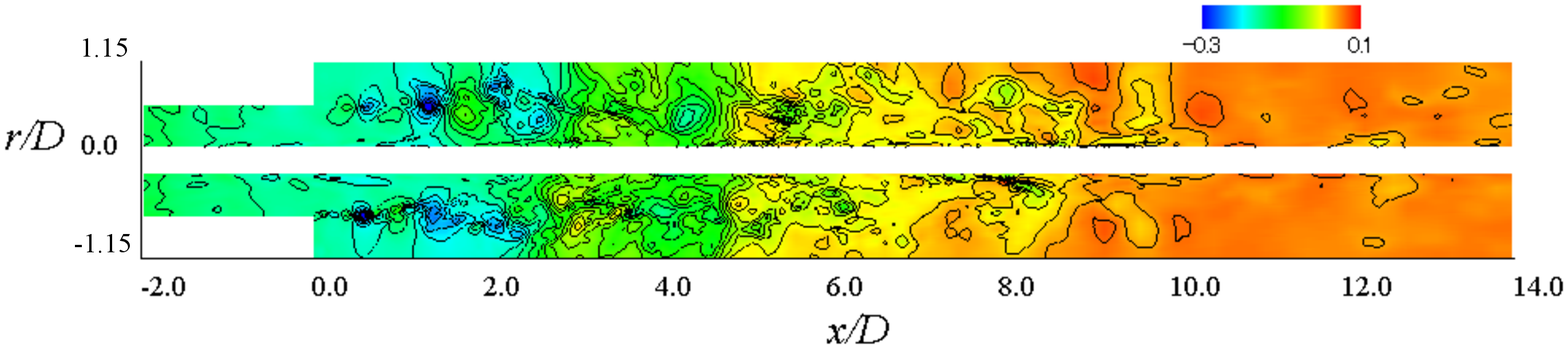} \\
(b) Pressure contours: Contour interval is 0.02 from $-3.0$ to 0.1. \\
\includegraphics[trim=0mm 5mm 0mm 10mm, clip, width=80mm]{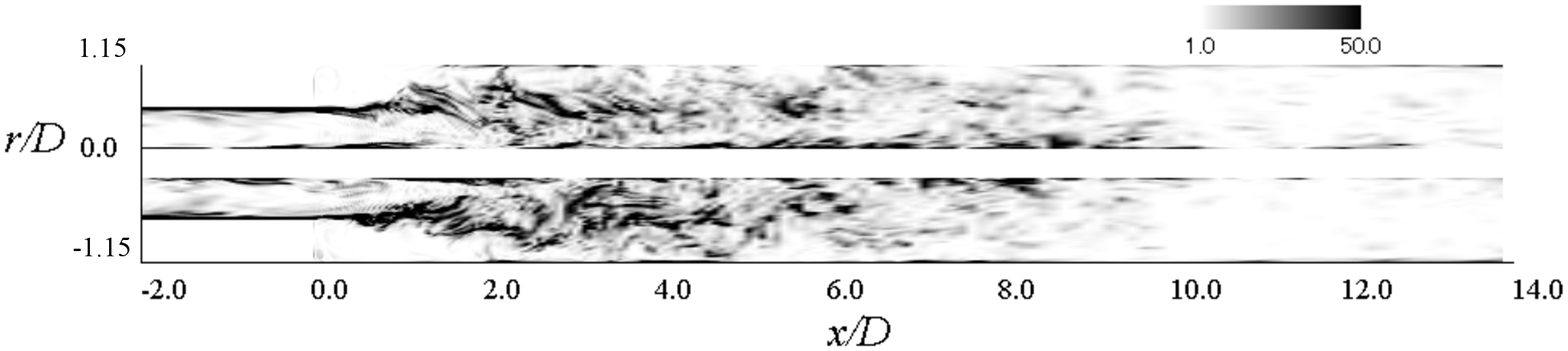} \\
(c) Enstrophy contours
\end{center}
\caption{Instantaneous flow field in $x$-$r$ plane for $\alpha=0.23$.}
\label{inst_r015}
\end{figure}

Figures \ref{inst_r05} to \ref{inst_r015} show the streamwise velocity, 
pressure, and enstrophy distributions in the $x$-$r$ plane on the central axis 
for each $\alpha$ value. 
As seen in the streamwise velocity distribution, 
an asymmetric recirculation region is formed on the outer pipe side 
by the shear layer separated at the step. 
The high-speed fluid decelerates downstream and diffuses 
throughout the flow channel. 
The smaller $\alpha$, the longer the recirculation region becomes 
toward the downstream.

In the pressure distribution, just downstream of the sudden expansion part, 
the pressure drops due to the vortex ring generated by the rollup of the shear layer. 
We find that around the center of the shear layer at $(r-r_1)/D=0.5$, 
high- and low-pressure changes are repeated alternately due to the vortex ring. 
Downstream, the vortex ring collapses, so the pressure change 
due to this vortex ring also disappears. 
The pressure is recovered in the redevelopment region 
downstream of the reattachment point. 
However, the small-scale vortex structure formed by the collapse of the vortex ring 
causes a local pressure drop mainly on the inner pipe side. 
The smaller $\alpha$ is, the more pronounced this pressure drop.

It can be seen from the enstrophy distribution that the flow separates 
at the step, and the separated shear layer extends downstream. 
The shear layer becomes unstable downstream 
and rolls up into a vortex, partly colliding with the wall surface 
on the outer pipe side, 
as seen near $x/D=2.0$ in the distribution of $\alpha=0.5$.

\begin{figure}[!t]
\begin{minipage}{0.33\linewidth}
\begin{center}
\includegraphics[trim=0mm 5mm 0mm 12mm, clip, width=50mm]{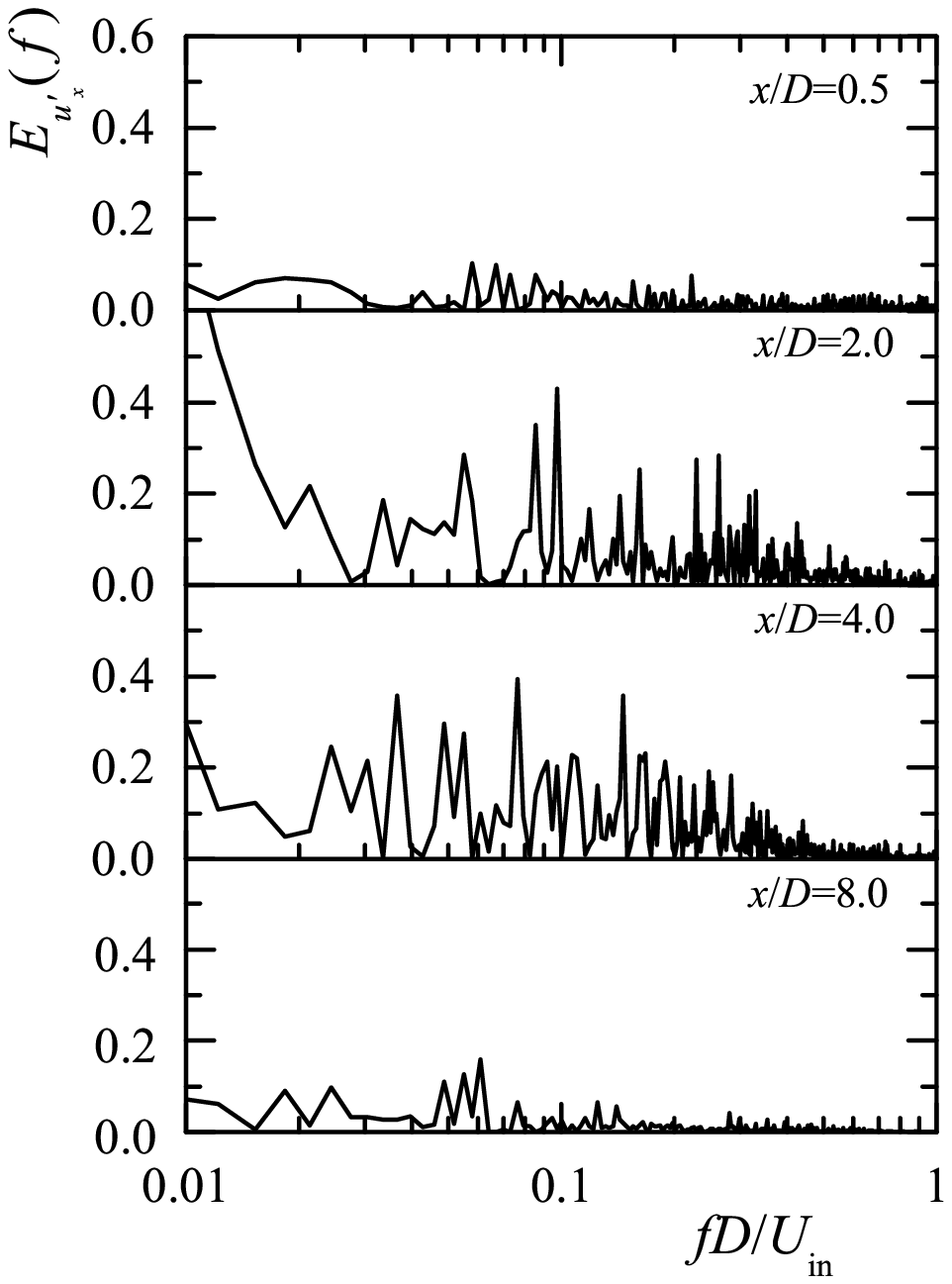} \\
(a) $\alpha=0.5$
\end{center}
\end{minipage}
\begin{minipage}{0.33\linewidth}
\begin{center}
\includegraphics[trim=0mm 5mm 0mm 12mm, clip, width=50mm]{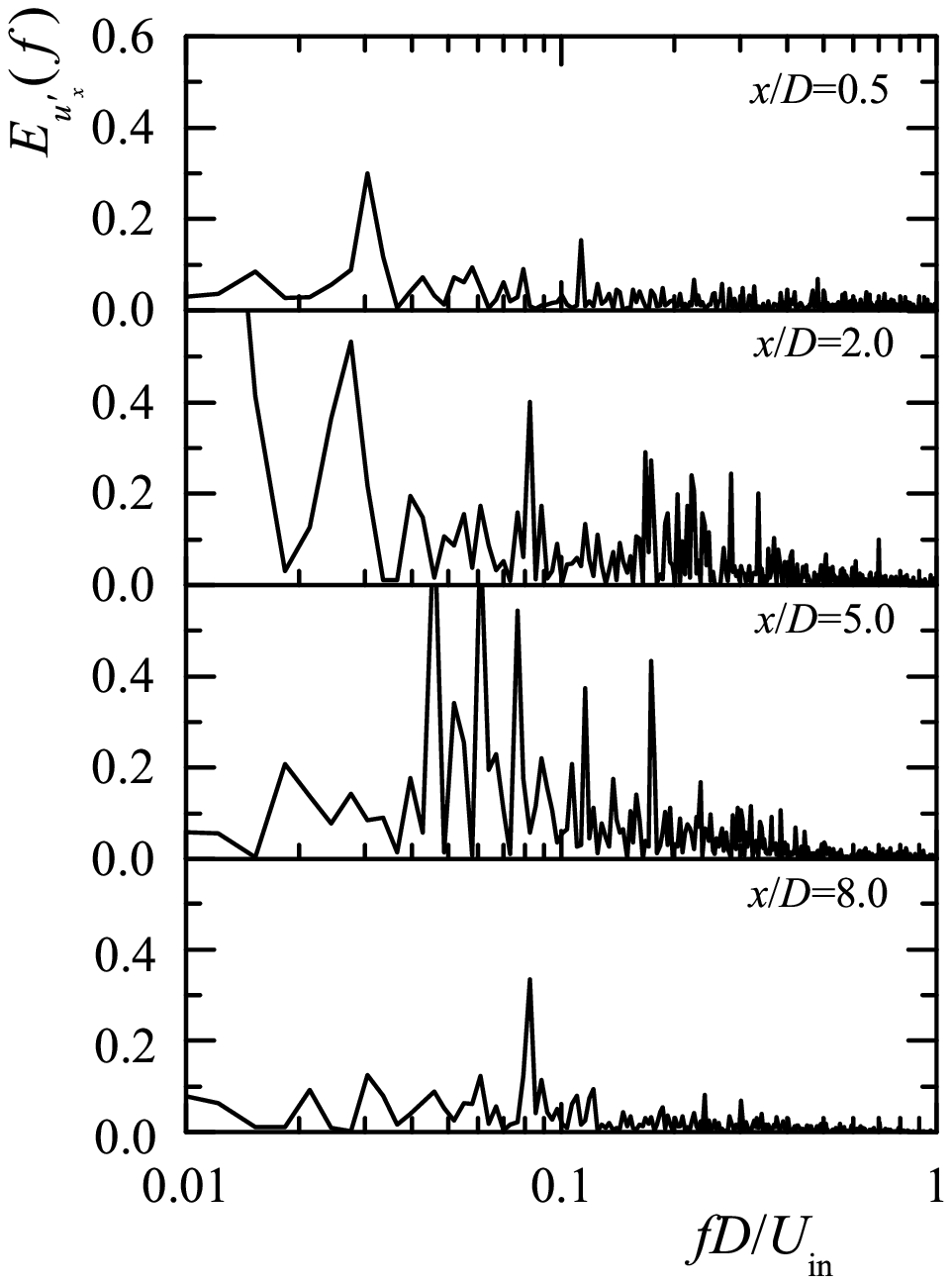} \\
(b) $\alpha=0.375$
\end{center}
\end{minipage}
\begin{minipage}{0.33\linewidth}
\begin{center}
\includegraphics[trim=0mm 5mm 0mm 12mm, clip, width=50mm]{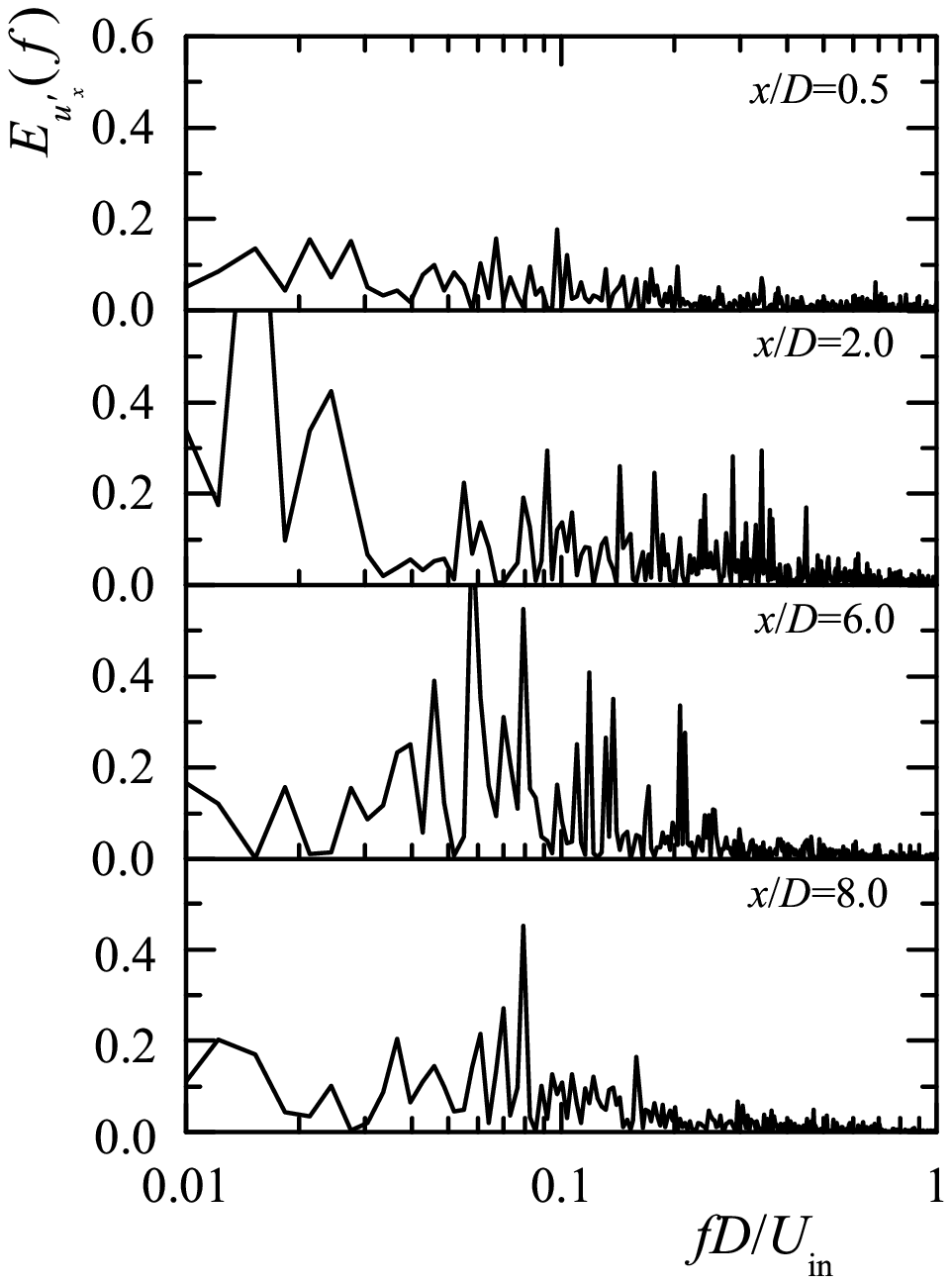} \\
(c) $\alpha=0.23$
\end{center}
\end{minipage}
\caption{Power spectrums of streamwise velocity fluctuation at $(r-r_1)/D=0.5$.}
\label{spctr}
\end{figure}

\subsection{Frequency characteristics}

Figure \ref{spctr} shows the power spectrum of the streamwise velocity fluctuation 
$u_x'$ at $(r-r_1)/D=0.5$ near the center of the separated shear layer. 
At $x/D=2.0$, remarkable peaks for $\alpha=0.5$, 0.375, and 0.23 occur 
at $f=0.097U_\mathrm{in}/D$, $0.083U_\mathrm{in}/D$, and $0.091U_\mathrm{in}/D$, respectively. 
This is the shedding frequency of the vortex ring. 
In addition, regardless of $\alpha$, several peaks can be seen 
from $f=0.07U_\mathrm{in}/D$ to $f=0.4U_\mathrm{in}/D$ due to the disturbance 
from the upstream. 
Furthermore, for $\alpha=0.5$, 0.375, and 0.23, peaks occur around 
$f=0.021U_\mathrm{in}/D$, $0.027U_\mathrm{in}/D$, and $0.024U_\mathrm{in}/D$, respectively, 
on the low frequency side. 
This is considered to represent the low frequency fluctuation of 
the separated shear layer. 
Many peaks due to many small-scale vortices caused by the collapse 
of the vortex ring appear in the distributions at $x/D=4.0$ for $\alpha=0.5$, 
$x/D=5.0$ for $\alpha=0.375$, and $x/D=6.0$ for $\alpha=0.23$.

\begin{figure}[!t]
\begin{minipage}{0.33\linewidth}
\begin{center}
\includegraphics[trim=0mm 5mm -5mm 7mm, clip, width=60mm]{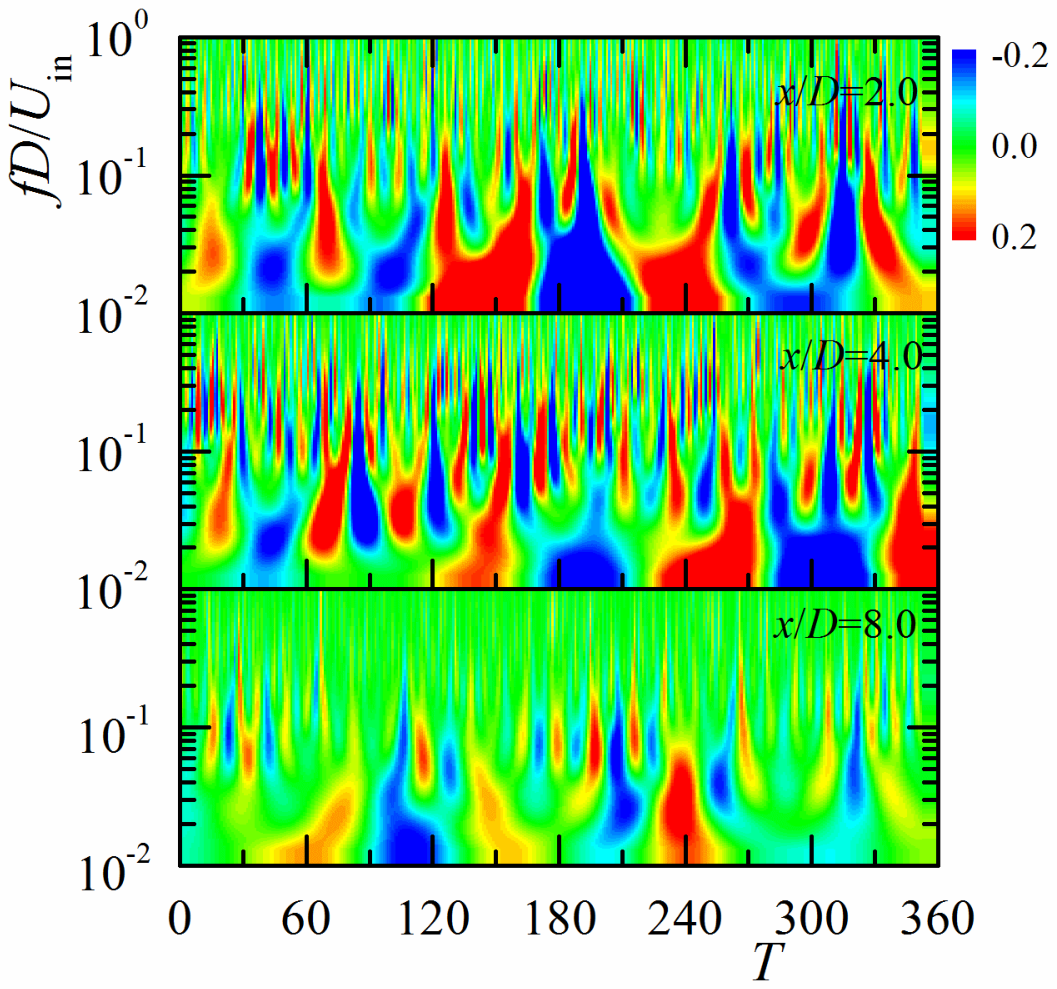} \\
(a) $\alpha=0.5$
\end{center}
\end{minipage}
\begin{minipage}{0.33\linewidth}
\begin{center}
\includegraphics[trim=0mm 5mm -5mm 7mm, clip, width=60mm]{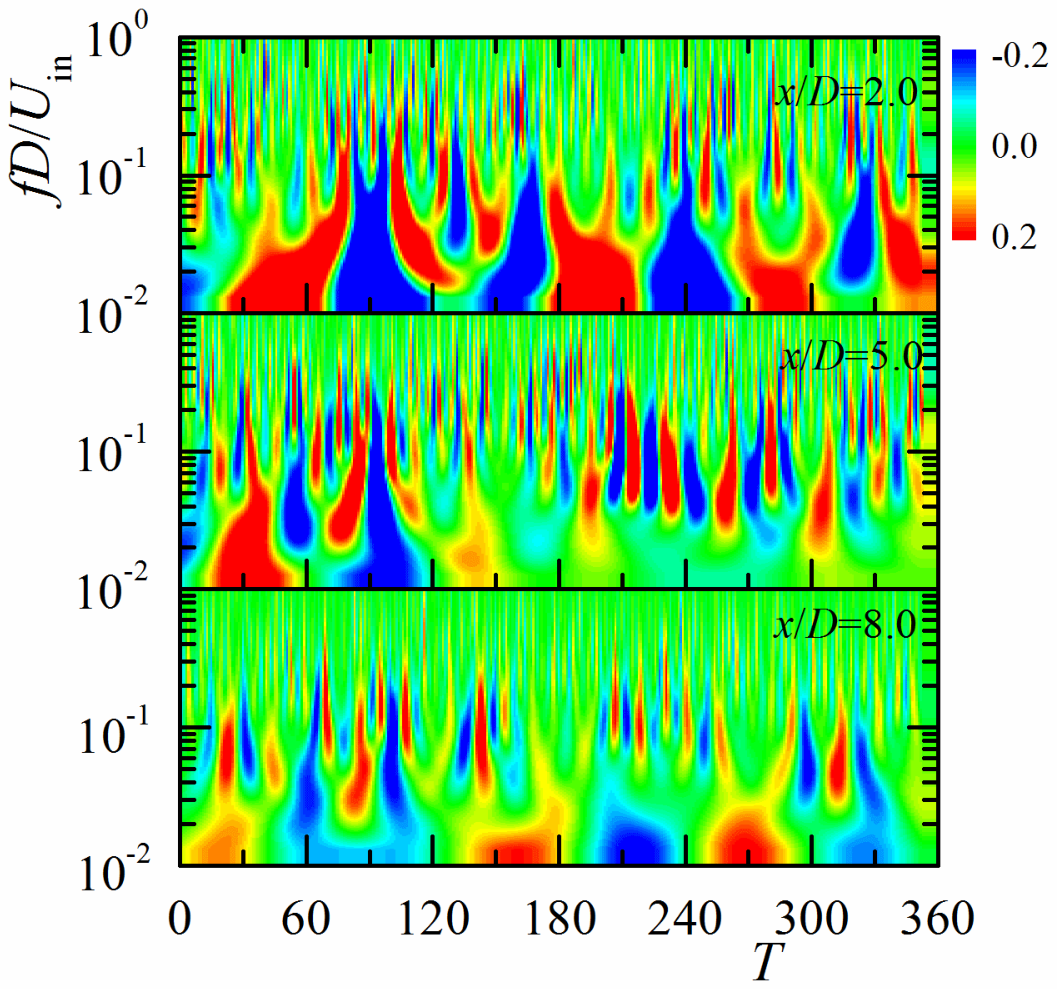} \\
(b) $\alpha=0.375$
\end{center}
\end{minipage}
\begin{minipage}{0.33\linewidth}
\begin{center}
\includegraphics[trim=0mm 5mm -5mm 7mm, clip, width=60mm]{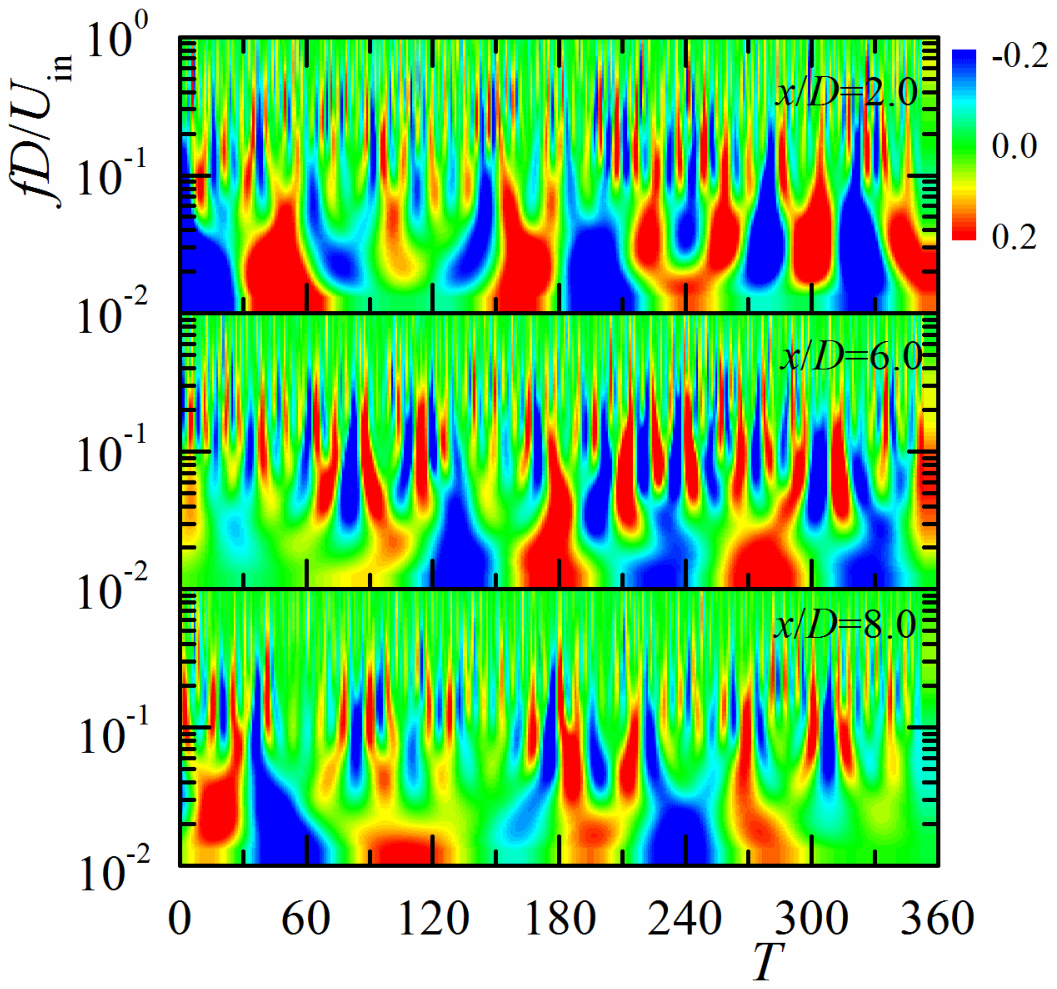} \\
(c) $\alpha=0.23$
\end{center}
\end{minipage}
\caption{Wavelet transform of streamwise velocity fluctuation at $(r-r_1)/D=0.5$.}
\label{wave}
\end{figure}

In this study, to clarify the spatiotemporal behavior of the vortex structure, 
we performed a wavelet transform, which is a kind of time-frequency analysis. 
The wavelet transform $W(a,b)$ is defined by the following equation 
by a signal $x(t)$ and the mother wavelet $\psi(t)$.
\begin{equation}
  W(a,b) = \int_{-\infty}^{\infty} x(t) \psi_{a,b}^{*}(t) dt
         = \frac{1}{\sqrt{a}} \int_{-\infty}^{\infty} x(t) \psi^{*} 
         \left( \frac{t - b}{a} \right) dt,
\end{equation}
where $a$ and $b$ are parameters for scaling and translating the mother wavelet. 
This study selected the Mexican hat defined by the following equation 
for the mother wavelet.
\begin{equation}
  \psi(t) = (1 - t^2) \exp \left( \frac{- t^2}{2} \right).
\end{equation}
This wavelet is a real wavelet that shows good temporal resolution 
at sufficient frequency resolution and is suitable for the analysis of 
spatiotemporal behavior of large-scale vortex structures 
\citep{Lee&Sung_2001}. 
Since the scaling parameter $a$ corresponds to the wavelet period, 
the relationship of $f \propto a^{-1}$ holds, 
and the relational expression between the scaling parameter and frequency 
in the Mexican hat is given as
\begin{equation}
  f = \frac{\sqrt{5/2}}{2 \pi a}.
\end{equation}
Figure \ref{wave} shows the wavelet transform obtained by streamwise velocity fluctuation 
at $(r-r_1)/D=0.5$. 
In the distribution of $x/D=2.0$, a striped pattern centered on 
$f=0.1U_\mathrm{in}/D$ at each $\alpha$ value appears periodically. 
This pattern represents the shedding of a vortex ring. In addition, 
low-frequency fluctuation of the separated shear layer occurs at the center of 
$f=0.03U_\mathrm{in}/D$ regardless of $\alpha$. 
Many striped patterns in the frequency band indicating 
the shedding of the vortex ring are present in the same time zone 
as the frequency band indicating the low-frequency fluctuation. 
Therefore, it is considered that the low-frequency fluctuation promotes 
the destabilization of the separated shear layer and the rollup of the shear layer 
into the vortex ring. 
In the distribution around the recirculation region, 
stripes appear periodically at a higher frequency 
than the shedding frequency of the vortex ring, regardless of $\alpha$. 
At these positions, small-scale vortex groups generated by the collapse of 
the vortex ring were confirmed, 
so it is considered that the striped pattern on the high-frequency side 
is due to the small-scale vortices. 
Since most of the striped patterns due to this vortex group exist 
at the same time zone as the striped pattern showing low-frequency fluctuation, 
it is considered that the vortex group is strongly affected 
by the low-frequency fluctuation. 
At $x/D=8.0$ downstream from the reattachment point, 
the smaller the $\alpha$, the stronger the striped pattern 
due to small-scale vortices and low-frequency fluctuations. 
Therefore, the smaller the $\alpha$ is, the more the effects of vortex groups 
and low-frequency fluctuations on the flow field appear further downstream.

\begin{figure}[!t]
\begin{minipage}{0.49\linewidth}
\begin{center}
\includegraphics[trim=3mm 0mm 0mm 0mm, clip, width=80mm]{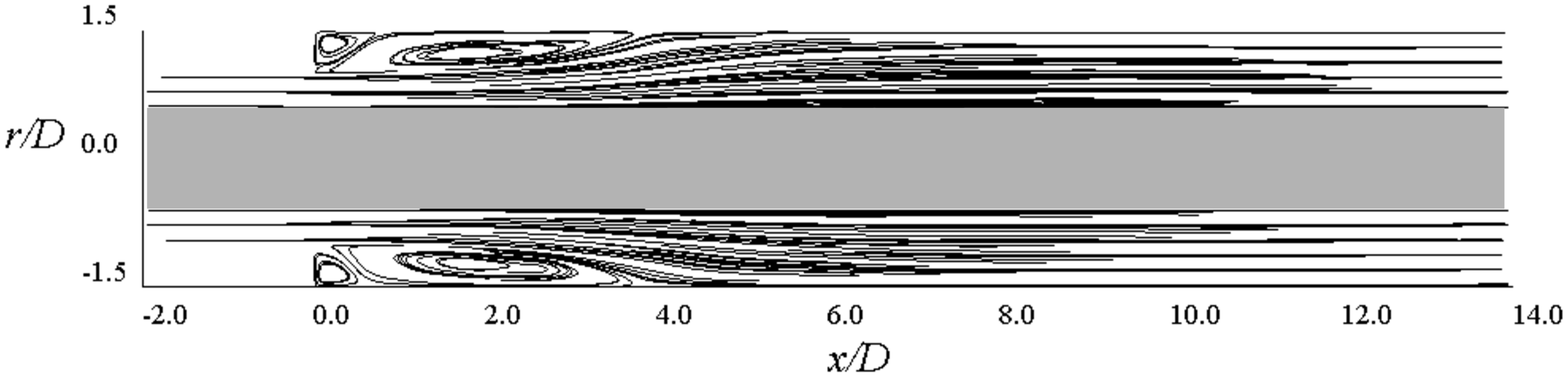} \\
(a) Streamlines \\
\includegraphics[trim=0mm 0mm 0mm 0mm, clip, width=80mm]{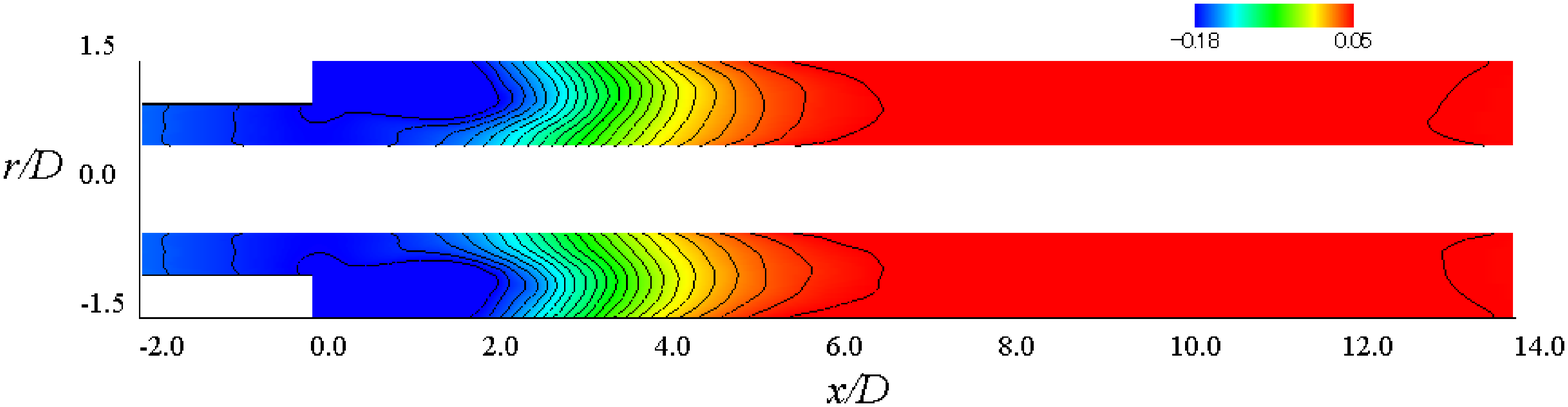} \\
(b) Pressure contours: Contour interval is 0.01 from $-0.18$ to 0.05. \\
\includegraphics[trim=0mm 0mm 0mm 0mm, clip, width=80mm]{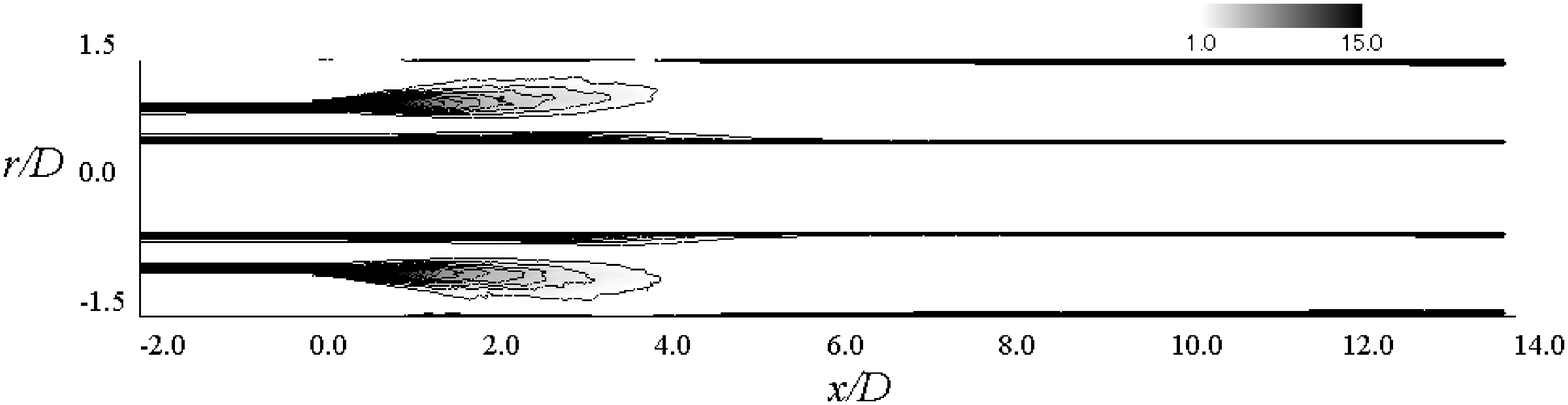} \\
(c) Enstorophy contours: Contour interval is 1.0 from 1.0 to 15.0.
\caption{Time--averaged flow field in $x-r$ plane for $\alpha=0.5$.}
\label{mean_r05}
\end{center}
\end{minipage}
%
%
\hspace*{0.02\linewidth}
\begin{minipage}{0.49\linewidth}
\begin{center}
\includegraphics[trim=3mm 0mm 0mm 0mm, clip, width=80mm]{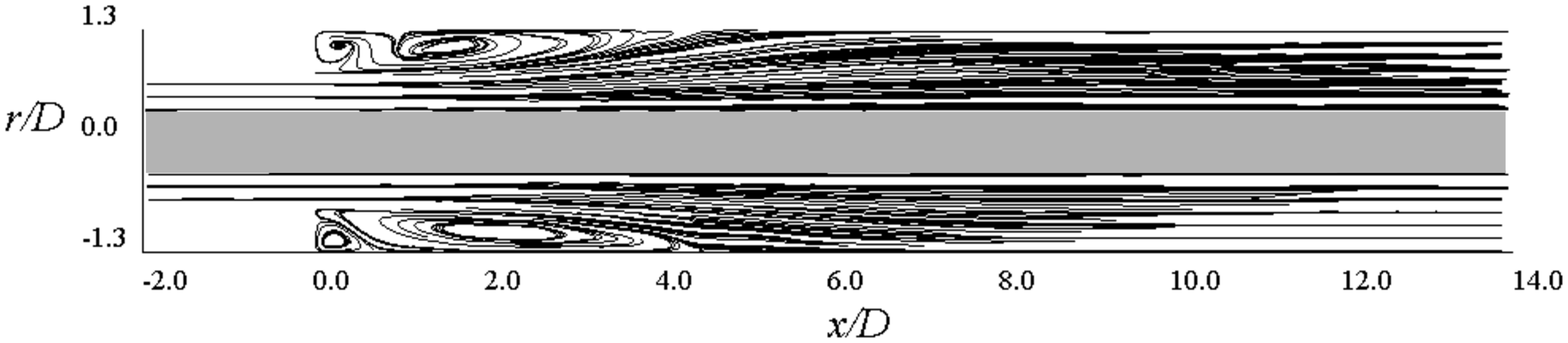} \\
(a) Streamlines \\
\includegraphics[trim=0mm 0mm 0mm 0mm, clip, width=80mm]{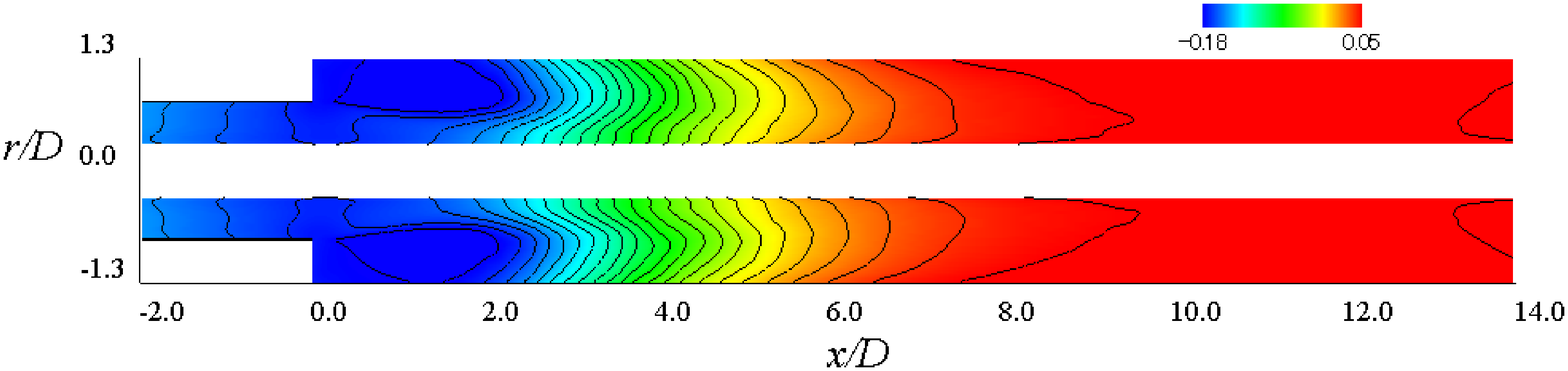} \\
(b) Pressure contours: Contour interval is 0.01 from $-0.18$ to 0.05. \\
\includegraphics[trim=0mm 0mm 0mm 0mm, clip, width=80mm]{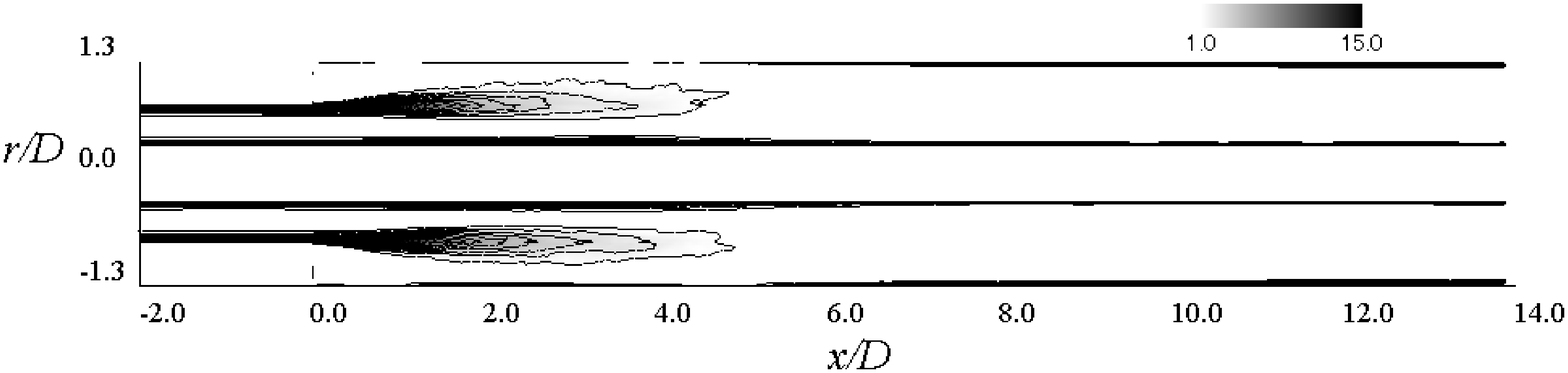} \\
(c) Enstorophy contours: Contour interval is 1.0 from 1.0 to 15.0.
\caption{Time--averaged flow field in $x-r$ plane for $\alpha=0.375$.}
\label{mean_r03}
\end{center}
\end{minipage}
%

%
\begin{center}
\includegraphics[trim=3mm 5mm 0mm 0mm, clip, width=80mm]{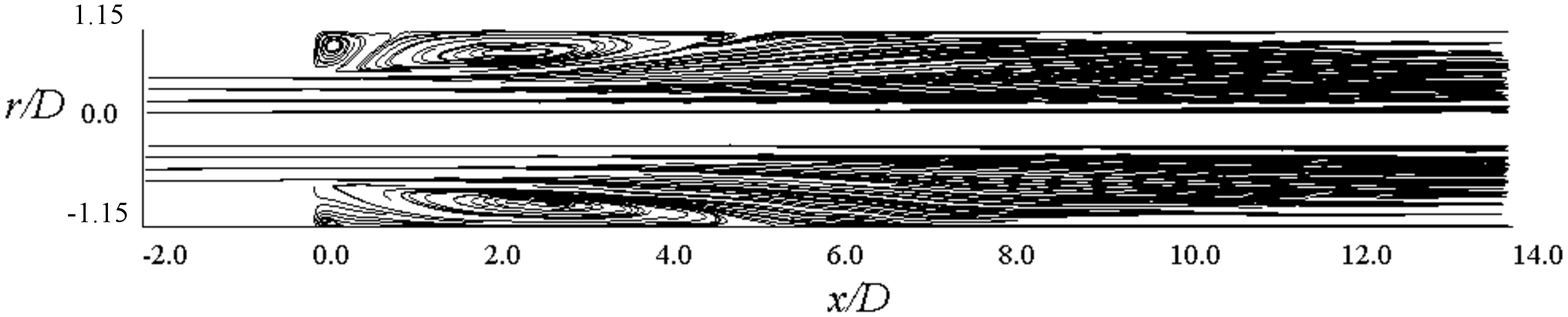} \\
(a) Streamlines \\
\includegraphics[trim=0mm 5mm 0mm 0mm, clip, width=80mm]{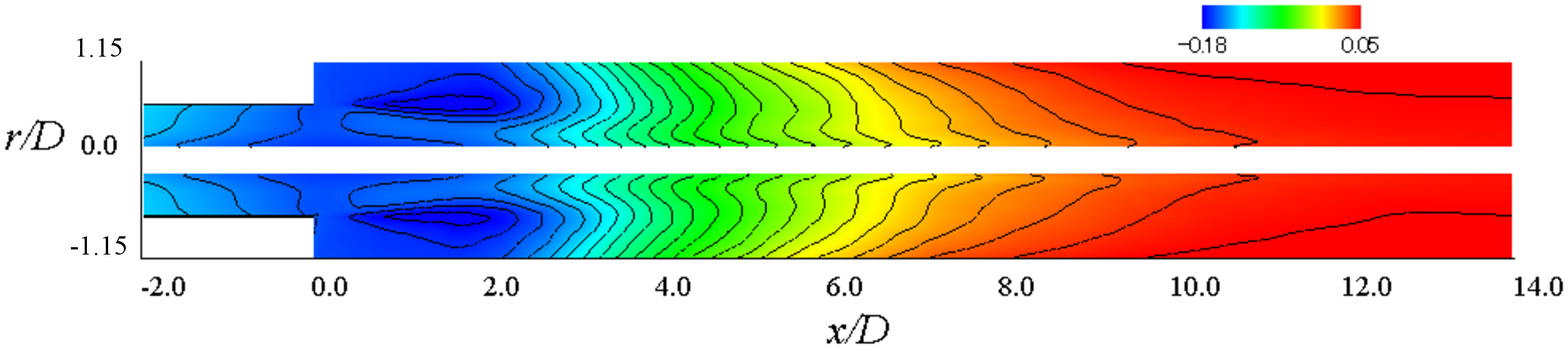} \\
(b) Pressure contours: Contour interval is 0.01 from $-0.18$ to 0.05. \\
\includegraphics[trim=0mm 5mm 0mm 0mm, clip, width=80mm]{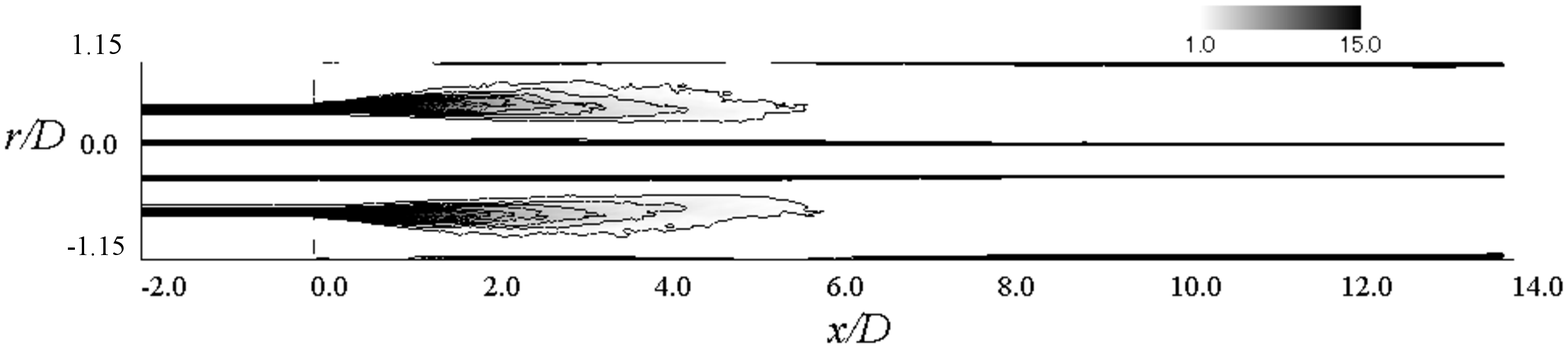} \\
(c) Enstorophy contours: Contour interval is 1.0 from 1.0 to 15.0. \\
\end{center}
\caption{Time--averaged flow field in $x-r$ plane for $\alpha=0.23$.}
\label{mean_r015}
\end{figure}

\begin{figure}[!t]
\centering
\begin{minipage}{0.49\linewidth}
\begin{center}
\includegraphics[trim=0mm 7mm 0mm 5mm, clip, width=75mm]{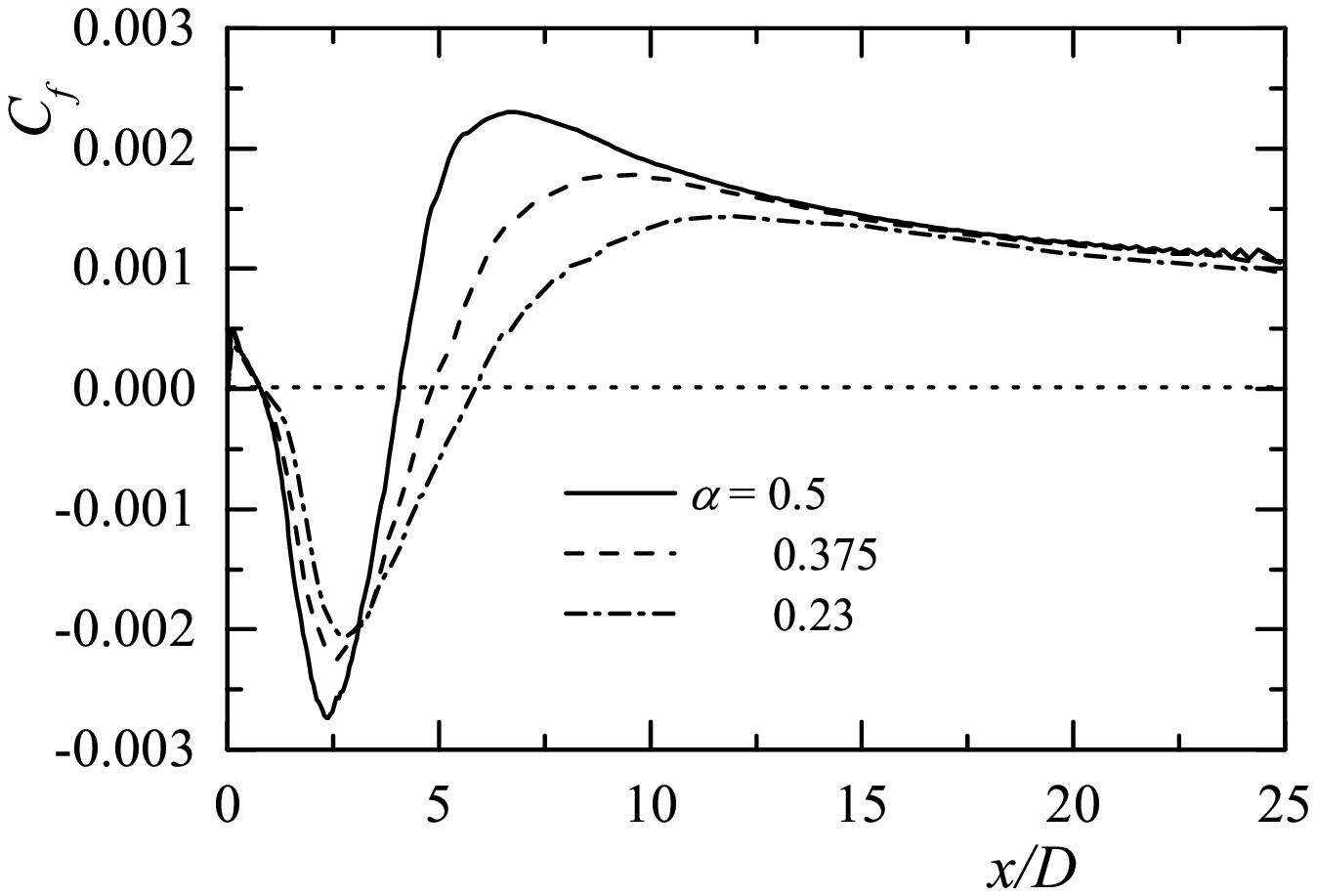} \\
(a) $C_f$
\end{center}
\end{minipage}
\begin{minipage}{0.49\linewidth}
\begin{center}
\includegraphics[trim=0mm 7mm 0mm 5mm, clip, width=75mm]{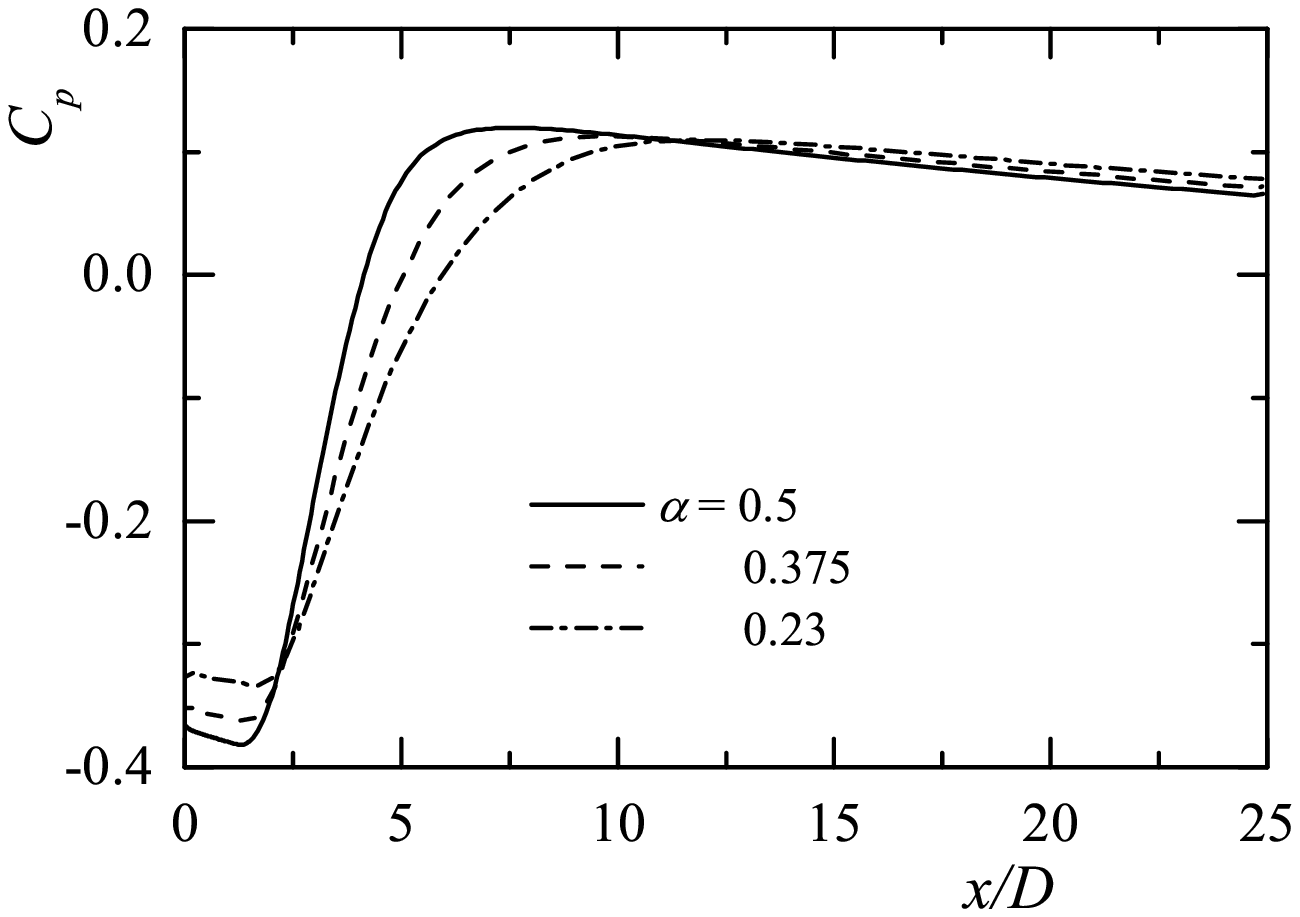} \\
(b) $C_p$
\end{center}
\end{minipage}
\caption{Time--averaged distributions of surface friction coefficient 
and surface pressure coefficient: 
---, $\alpha=0.5$; \mbox{- - -, $\alpha=0.375$}; - $\cdot$ -, $\alpha=0.23$.}
\label{cf_cp}
\end{figure}

\begin{figure}[!t]
\begin{minipage}{0.49\linewidth}
\begin{center}
\includegraphics[trim=0mm 7mm 0mm 5mm, clip, width=80mm]{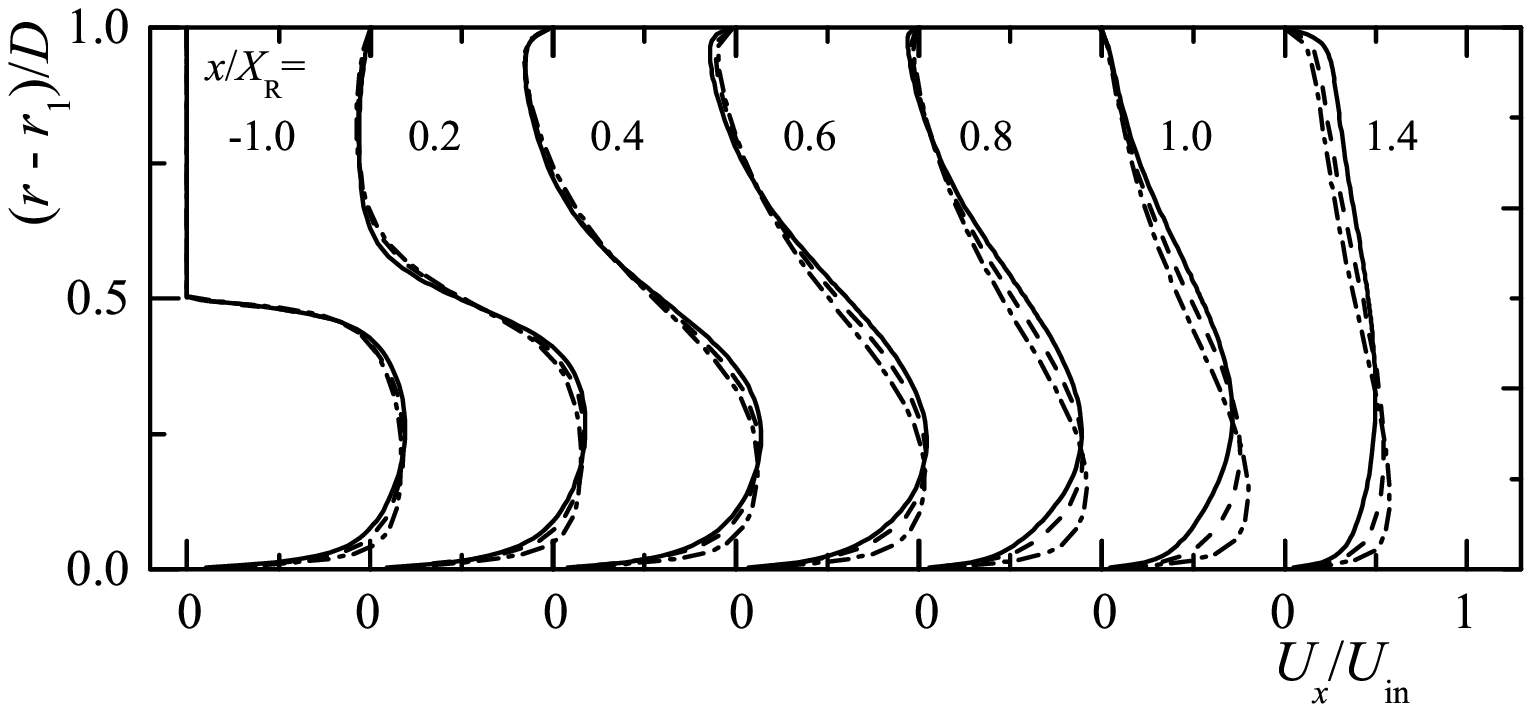} \\
(a) $U_x/U_\mathrm{in}$
\end{center}
\end{minipage}
\begin{minipage}{0.49\linewidth}
\begin{center}
\includegraphics[trim=0mm 7mm 0mm 5mm, clip, width=80mm]{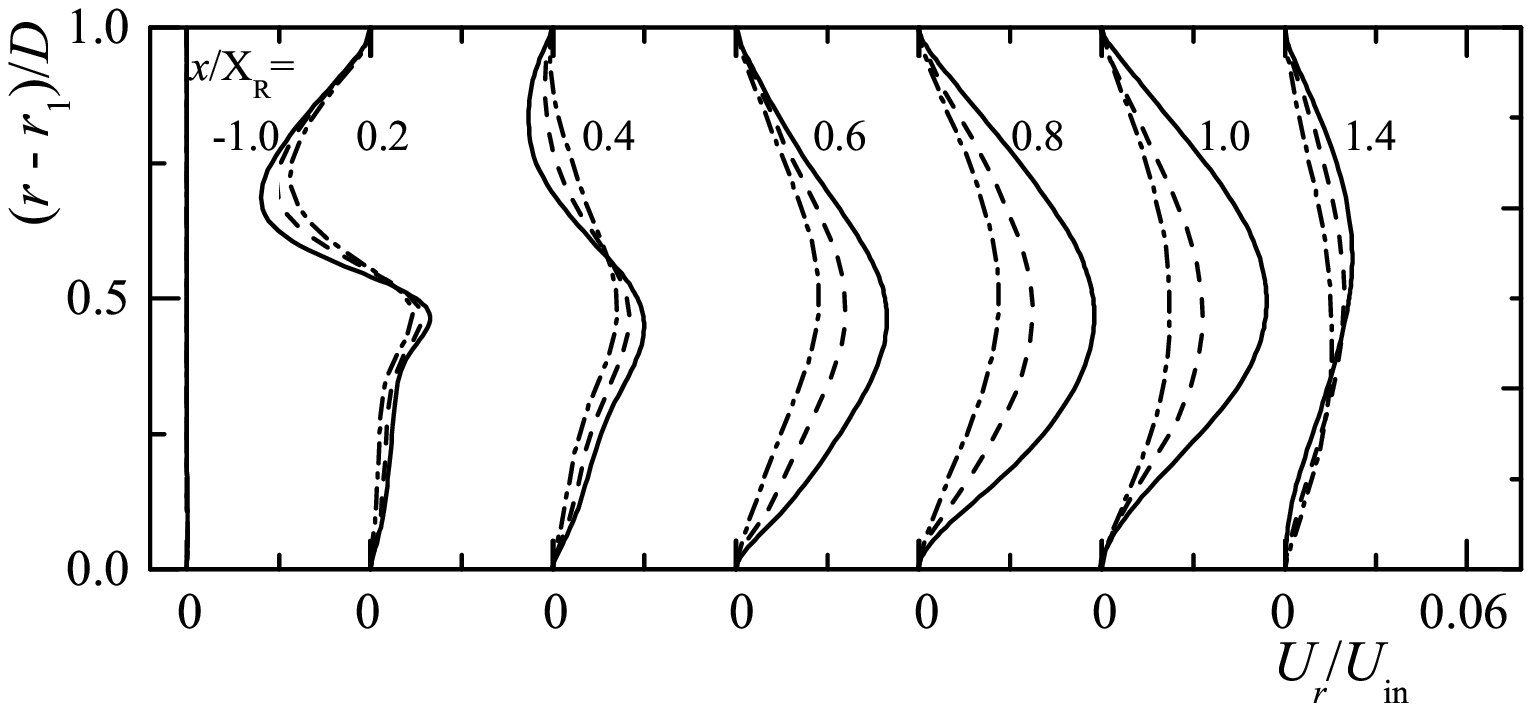} \\
(b) $U_r/U_\mathrm{in}$
\end{center}
\end{minipage}
\caption{Time--averaged velocity distributions: 
---, $\alpha=0.5$; - - -, $\alpha=0.375$; - $\cdot$ -, $\alpha=0.23$.}
\label{vel_mean}
\end{figure}

\subsection{Time-averaged flow field}

Figures \ref{mean_r05} to \ref{mean_r015} show the time-averaged streamline, 
pressure, and enstrophy distributions for each $\alpha$ value. 
In the streamline distribution, regardless of $\alpha$, 
the flow separated at the step attaches to the outer pipe wall surface downstream, 
generating a separation bubble. 
Furthermore, a secondary vortex is formed on the outer pipe side 
just downstream of the step.

The pressure drops sharply because a vortex ring is shed at the step. 
Downstream, the vortex ring collapses and splits into small vortices, 
so the pressure rises around $x/D=3.5$ for $\alpha=0.5$, 
$x/D=4.0$ for $\alpha=0.375$, and $x/D=4.5$ for $\alpha=0.23$. 
Because the small-scale vortices decay downstream of the reattachment point, 
for $\alpha=0.5$, 0.375, and 0.23, the pressure becomes almost constant 
downstream from about $x/D=6.0$, 8.0, and 10.0, respectively. 
In addition, there is a region where the pressure recovery on the inner pipe side 
is delayed compared to the outer pipe side. 
This is because the small-scale vortex group generated 
by the collapse of the vortex ring is concentrated on the inner pipe side downstream, 
and a tubular longitudinal vortex structure is formed near the wall surface. 
The smaller $\alpha$, the more pronounced this tendency.

In the enstrophy distribution, the smaller $\alpha$ is, 
the more downstream the separated shear layer extends. 
Regardless of $\alpha$, the shear layer diffuses in the radial direction downstream, 
and the shear of the velocity decays.

Figure \ref{cf_cp} shows the time-averaged distributions of surface friction coefficient $C_f$ 
and surface pressure coefficient $C_p$, 
which are averaged in the circumferential direction, 
on the outer pipe side. 
$C_f$ has a maximum value just downstream of the step 
because the recirculation of the flow generates a secondary vortex.
$C_f$ decreased sharply downstream due to the rollup of the shear layer, 
and for $\alpha=0.5$, 0.375, and 0.23, $C_f$ becomes minimum 
at $x/D=2.3$, 2.6, and 2.9, respectively. 
This minimum indicates a strong reverse flow. 
Further downstream, $C_f$ increases sharply, reaches a maximum 
and then approaches a constant value.
For $\alpha=0.5$, 0.375, and 0.23, $C_p$ becomes minimum 
at $x/D=1.3$, 1.36, and 1.65, respectively. 
Downstream, the pressure recovers sharply, and for $\alpha=0.5$, 0.375, 
and 0.23, $C_p$ gradually approaches a constant value 
around $x/D=7.0$, 9.0, and 11.0.

Figure \ref{vel_mean} shows the time-averaged streamwise 
and radial velocity distributions, 
which are averaged in the circumferential direction. 
In Fig. \ref{vel_mean} (a), reverse flow occurs on the outer pipe side 
downstream of the step, 
indicating the existence of a recirculation region. 
For all $\alpha$ values, the high-speed fluid ejected from the sudden expansion part 
decays downstream, 
and the velocity distribution becomes uniform. 
In addition, the velocity distribution for each $\alpha$ value 
just downstream of the step is almost the same near the inner pipe wall surface. 
However, the larger $\alpha$, the greater the deceleration 
in the vicinity of the inner pipe wall surface downstream. 
It is considered from this result that the larger $\alpha$, 
the stronger the influence of the inner pipe wall surface on the annular flow.

In Fig. \ref{vel_mean} (b), the radial velocity is almost zero 
in the pipe upstream of the step. 
At $x/X_\mathrm{R}=0.2$ and 0.4, there are minimums inside the recirculation region. 
Upstream from the reattachment point of $x/X_\mathrm{R} \leq 1.0$, 
maximums induced by the vortex ring exist around $(r-r_1)/D=0.5$. 
In addition, upstream of the reattachment point, 
the smaller $\alpha$, the slower the radial velocity.

\begin{figure}[!t]
\begin{minipage}{0.49\linewidth}
\begin{center}
\includegraphics[trim=2mm 3mm 0mm 0mm, clip, width=70mm]{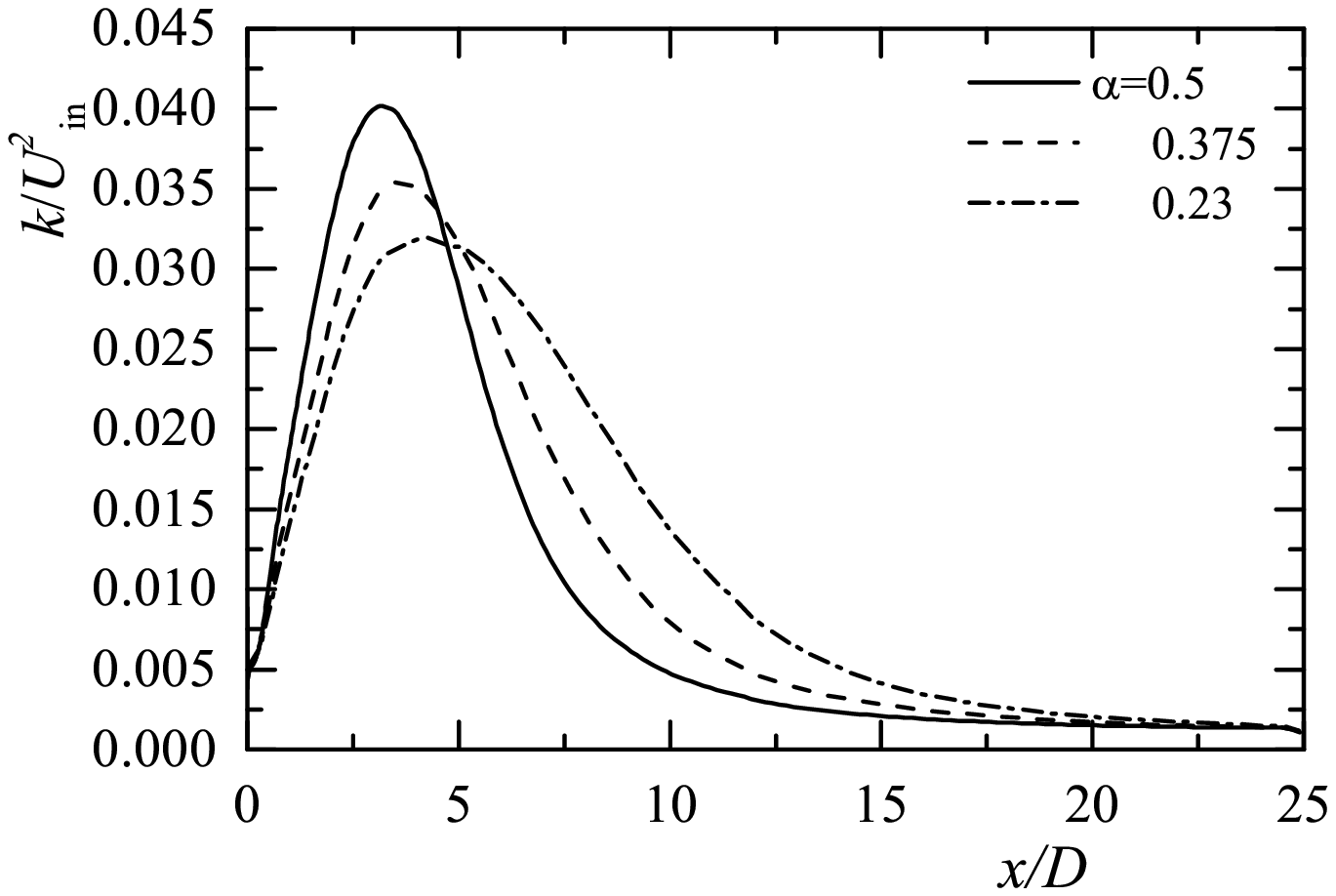} \\
(a) $k/U_\mathrm{in}^{2}$
\end{center}
\end{minipage}
\begin{minipage}{0.49\linewidth}
\begin{center}
\includegraphics[trim=2mm 3mm -3mm 0mm, clip, width=70mm]{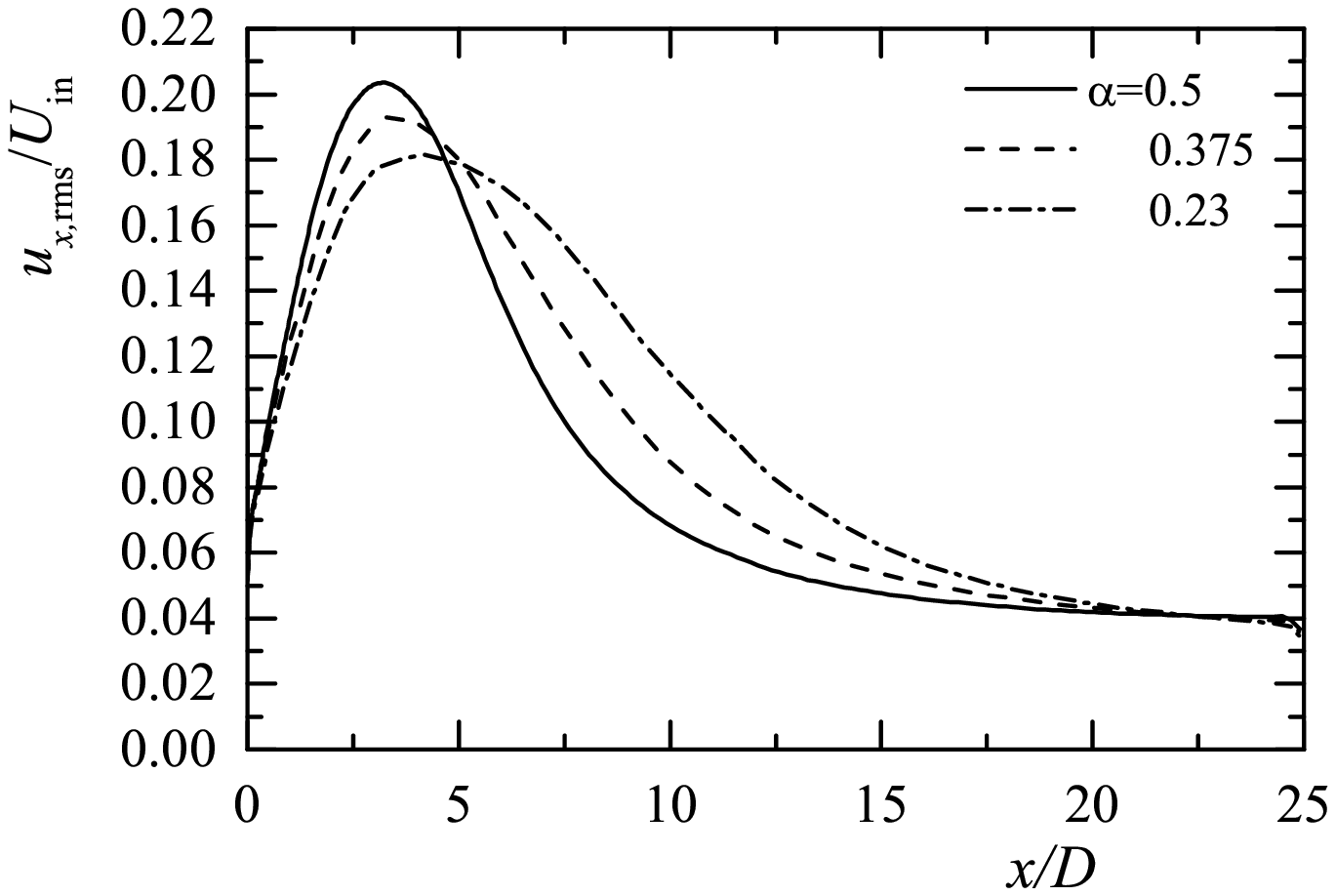} \\
(b) $u_{x,\mathrm{rms}}/U_\mathrm{in}$
\end{center}
\end{minipage}
\begin{minipage}{0.49\linewidth}
\begin{center}
\includegraphics[trim=0mm 3mm 0mm 0mm, clip, width=70mm]{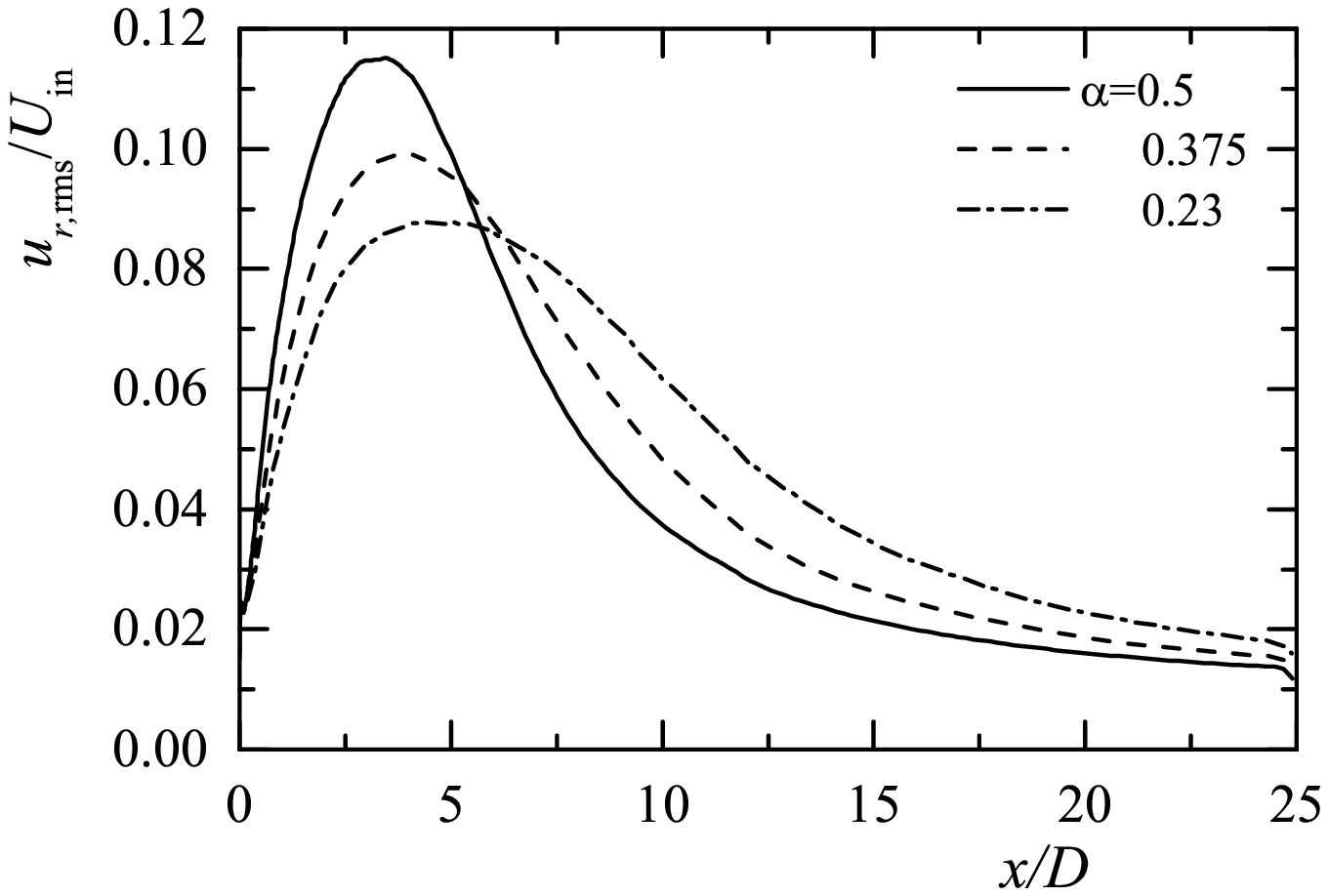} \\
(c) $u_{r,\mathrm{rms}}/U_\mathrm{in}$
\end{center}
\end{minipage}
\begin{minipage}{0.49\linewidth}
\begin{center}
\includegraphics[trim=0mm 3mm 0mm 0mm, clip, width=70mm]{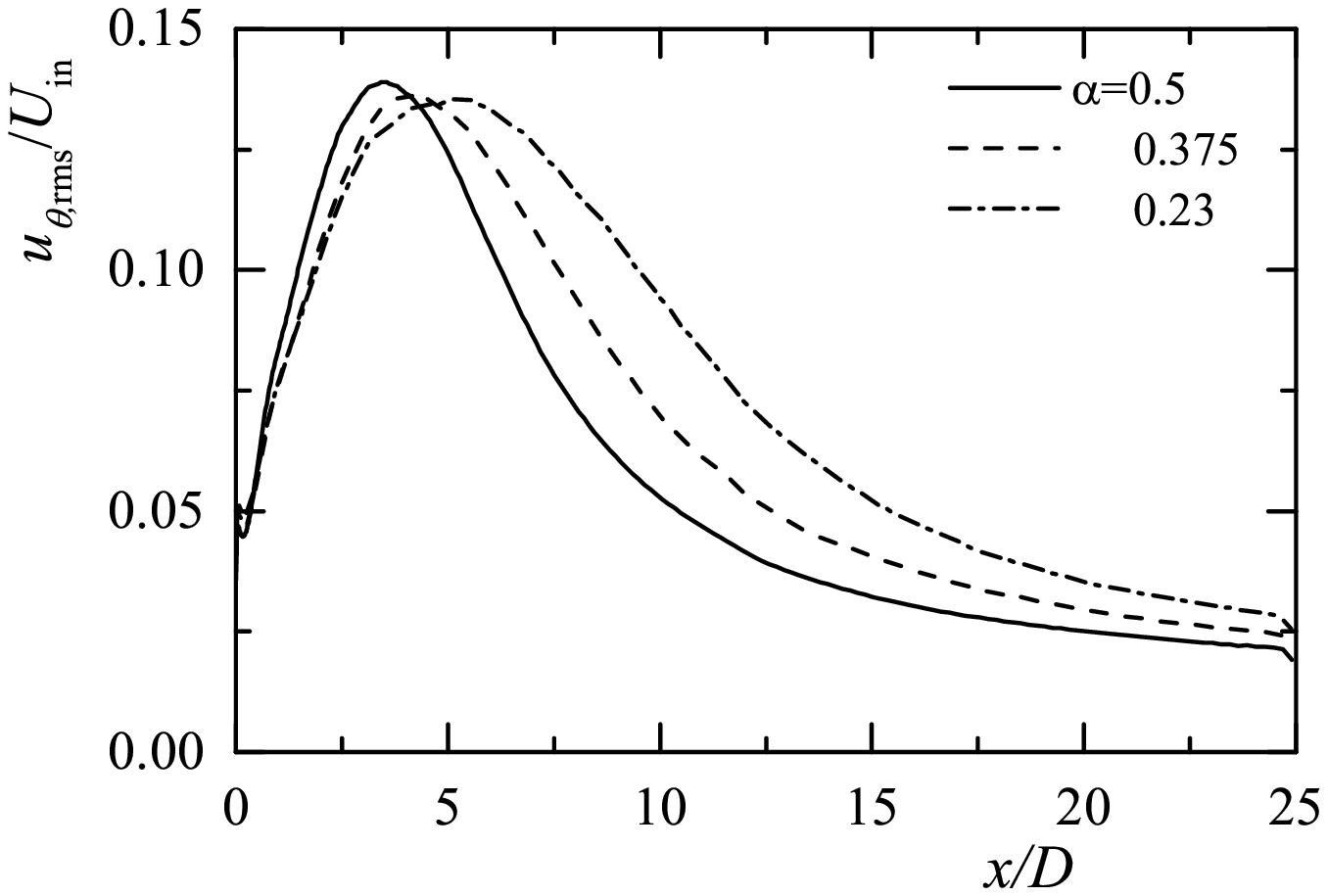} \\
(d) $u_{\theta,\mathrm{rms}}/U_\mathrm{in}$
\end{center}
\end{minipage}
\caption{Space--averaged distributions of turbulence kinetic energy and 
turbulence intensity of velocity fluctuation in $r-\theta$ plane.}
\label{rms_ave}
\end{figure}

\begin{figure}[!t]
\begin{minipage}{0.49\linewidth}
\begin{center}
\includegraphics[trim=0mm 11mm 0mm 3mm, clip, width=80mm]{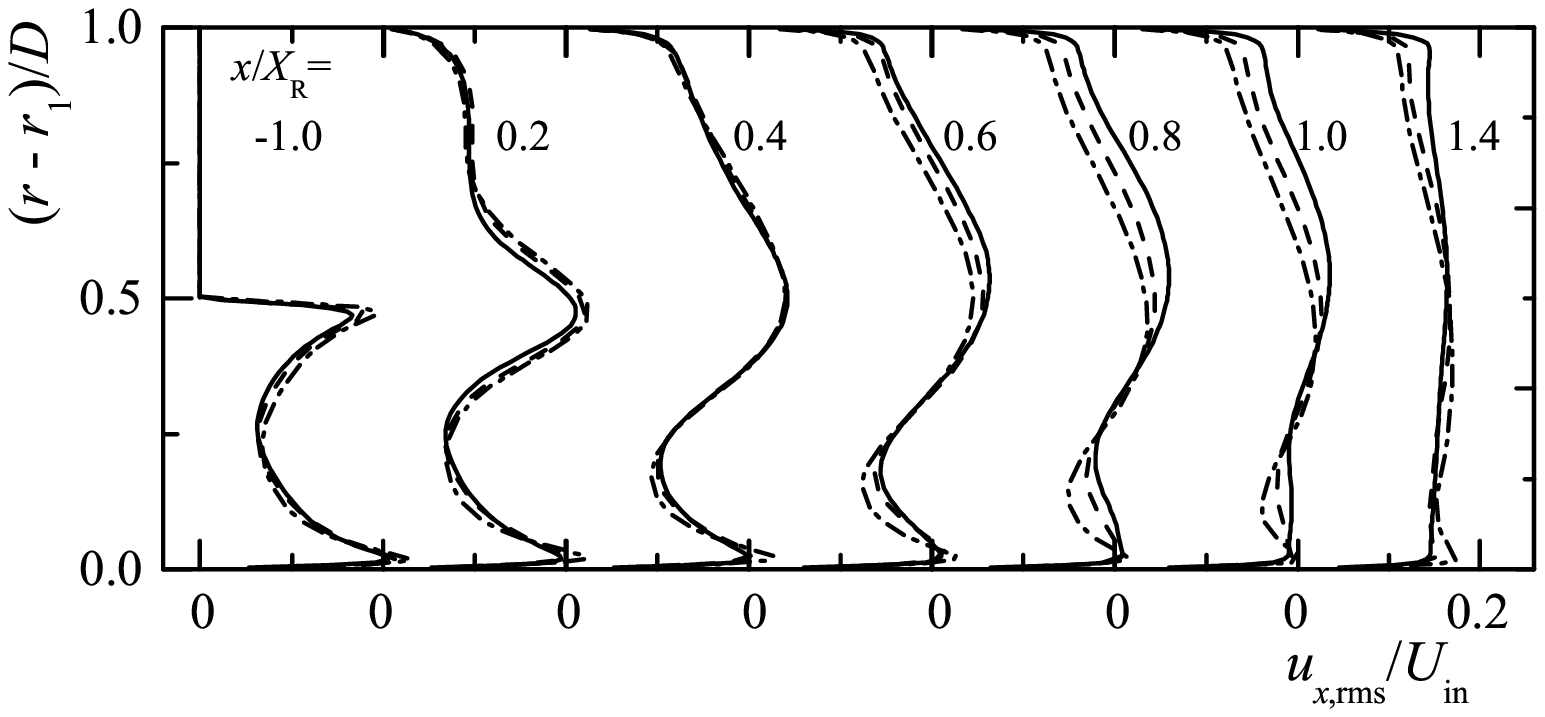} \\
(a) $u_{x,\mathrm{rms}}/U_\mathrm{in}$
\end{center}
\end{minipage}
\begin{minipage}{0.49\linewidth}
\begin{center}
\includegraphics[trim=0mm 11mm 0mm 3mm, clip, width=80mm]{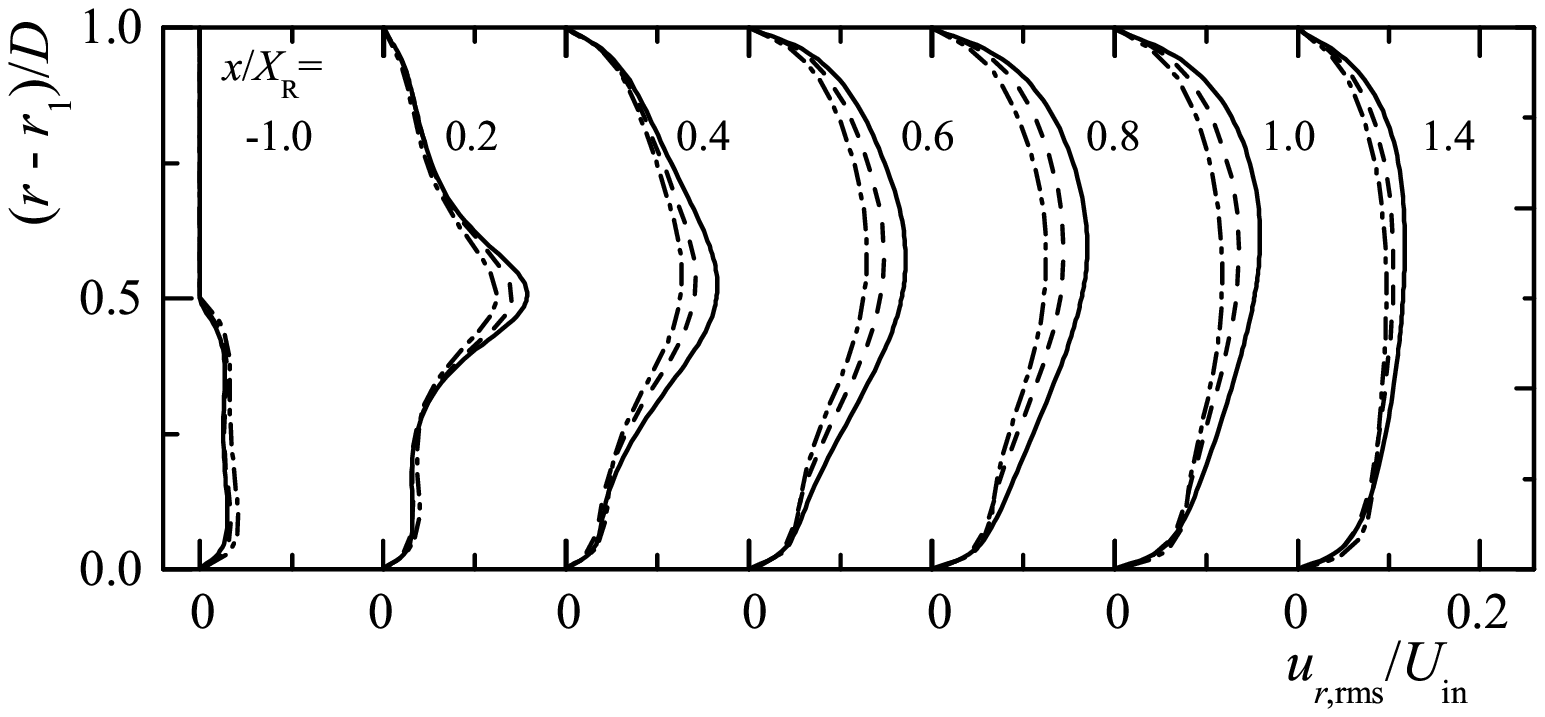} \\
(b) $u_{r,\mathrm{rms}}/U_\mathrm{in}$
\end{center}
\end{minipage}
\begin{minipage}{0.49\linewidth}
\begin{center}
\includegraphics[trim=0mm 11mm 0mm 3mm, clip, width=80mm]{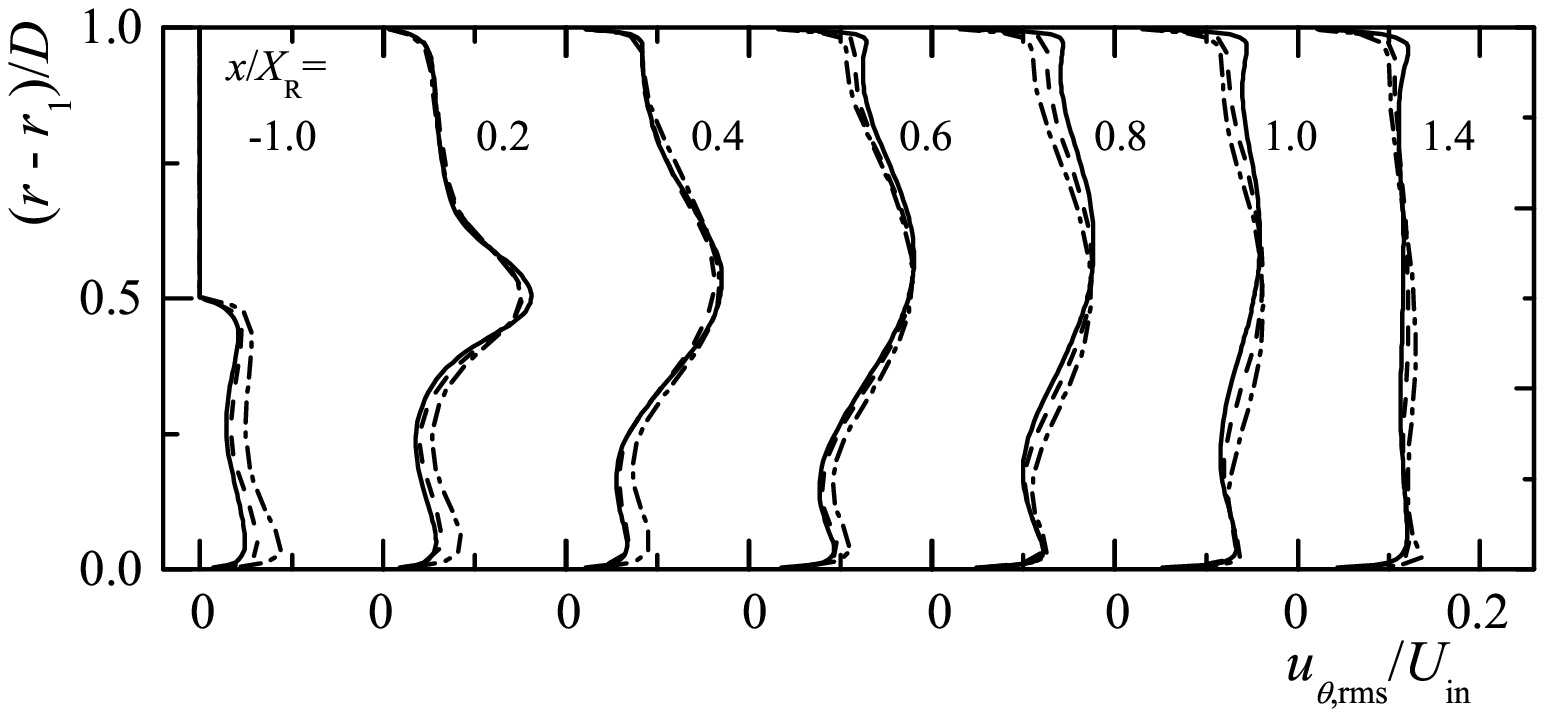} \\
(c) $u_{\theta,\mathrm{rms}}/U_\mathrm{in}$
\end{center}
\end{minipage}
\begin{minipage}{0.49\linewidth}
\begin{center}
\includegraphics[trim=0mm 11mm 0mm 3mm, clip, width=80mm]{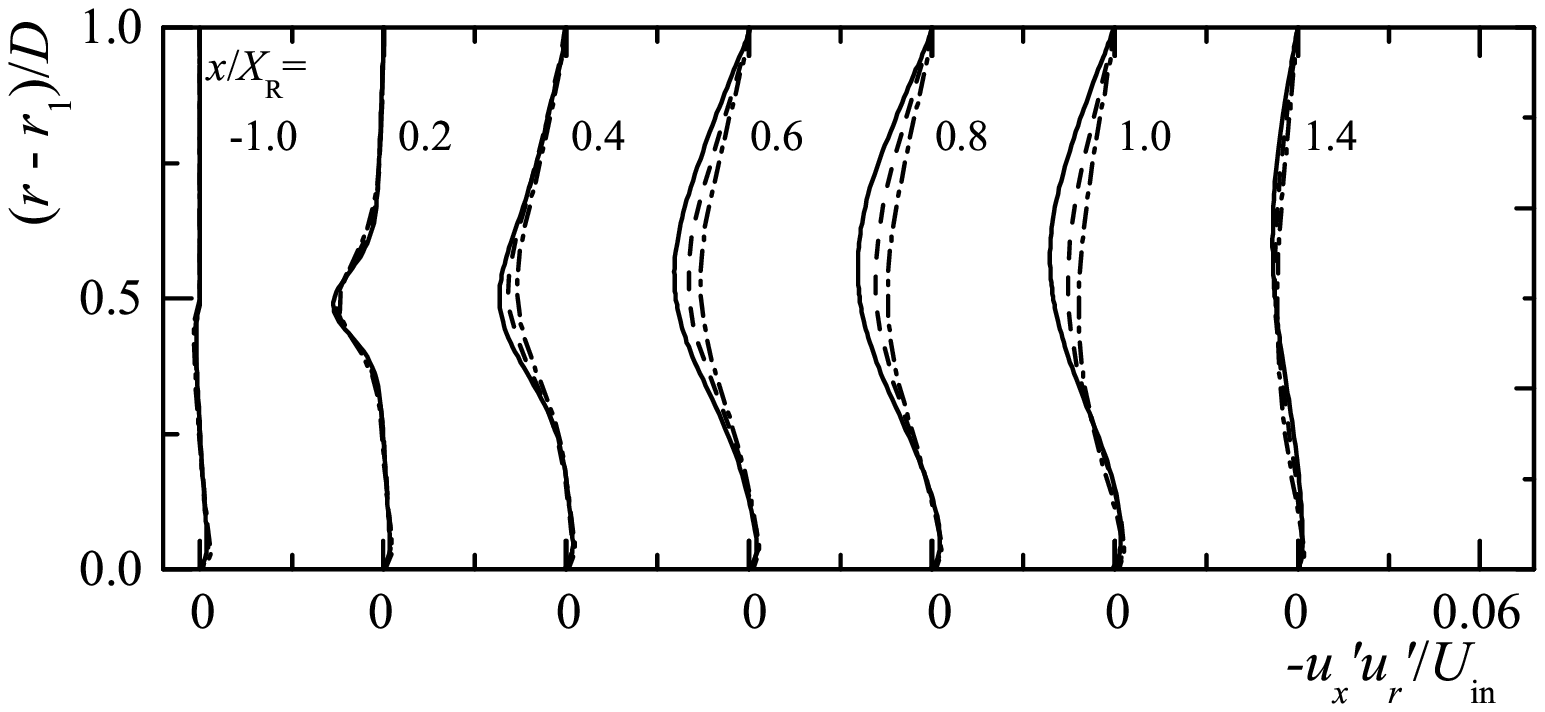} \\
(d) ${-u_x'}{u_r'}$
\end{center}
\end{minipage}
\caption{Turbulence intensity distributions of velocity fluctuations 
and Reynolds shear stress distributions: 
\mbox{---, $\alpha=0.5$}; - - -, $\alpha=0.375$; - $\cdot$ -, $\alpha=0.23$.}
\label{rms}
\end{figure}

\begin{figure}[!t]
\centering
\includegraphics[trim=0mm 11mm 0mm 3mm, clip, width=80mm]{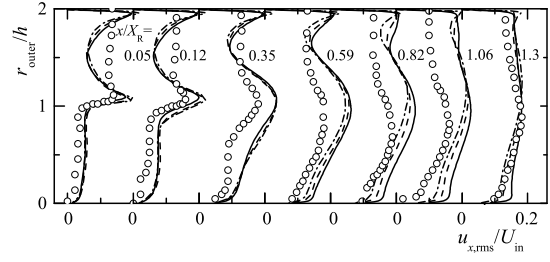} \\
\caption{Turbulence intensity distributions of velocity fluctuations: 
---, $\alpha=0.5$; - - -, $\alpha=0.375$; - $\cdot$ -, $\alpha=0.23$; 
\mbox{$\circ$, Yanhua et al. ($Re_h=3450$, $ER=1.01$)}.}
\label{urms_datou}
\end{figure}

\begin{figure}[t]
\begin{minipage}{0.49\linewidth}
\begin{center}
\includegraphics[trim=0mm 7mm 0mm 0mm, clip, width=80mm]{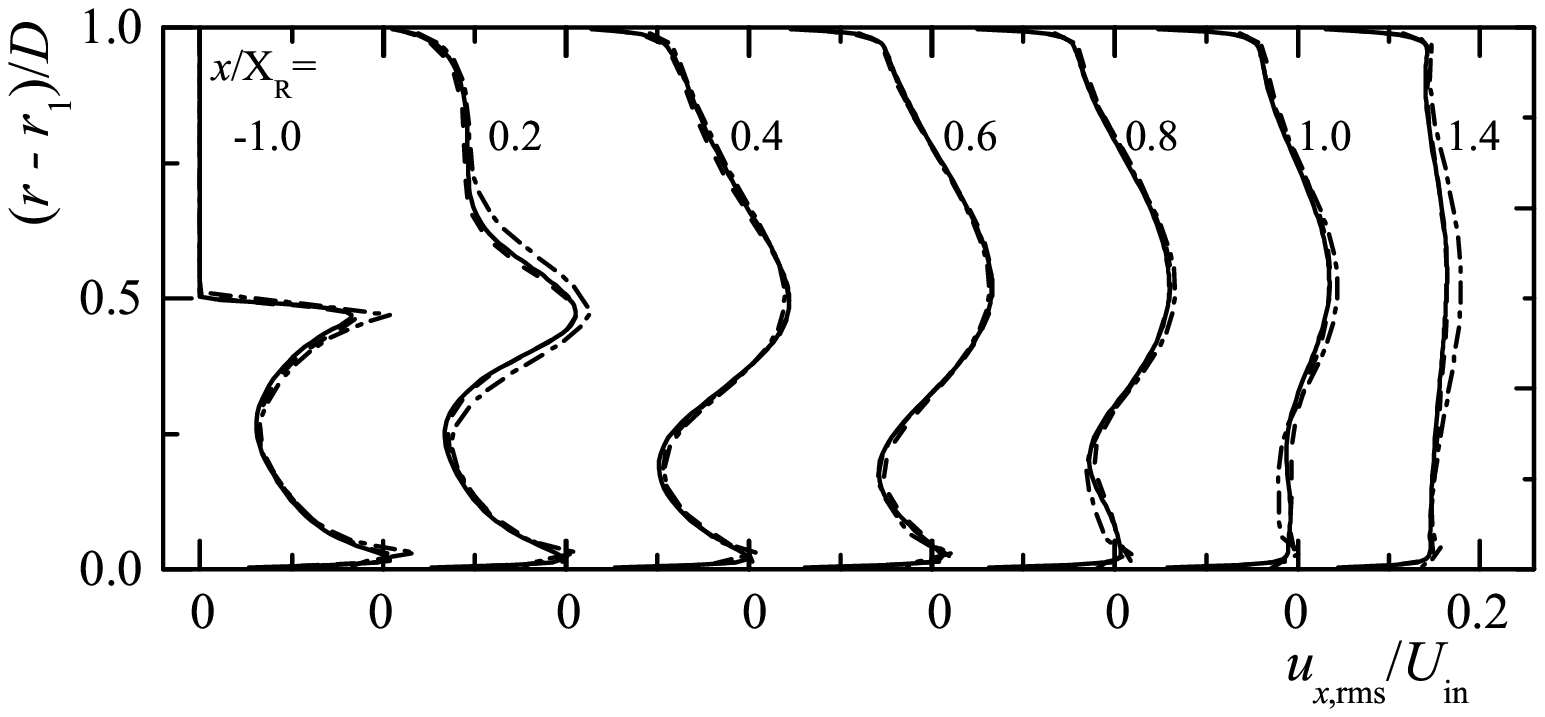} \\
(a) $u_{x,\mathrm{rms}}/U_\mathrm{in}$
\end{center}
\end{minipage}
\begin{minipage}{0.49\linewidth}
\begin{center}
\includegraphics[trim=0mm 7mm 0mm 0mm, clip, width=80mm]{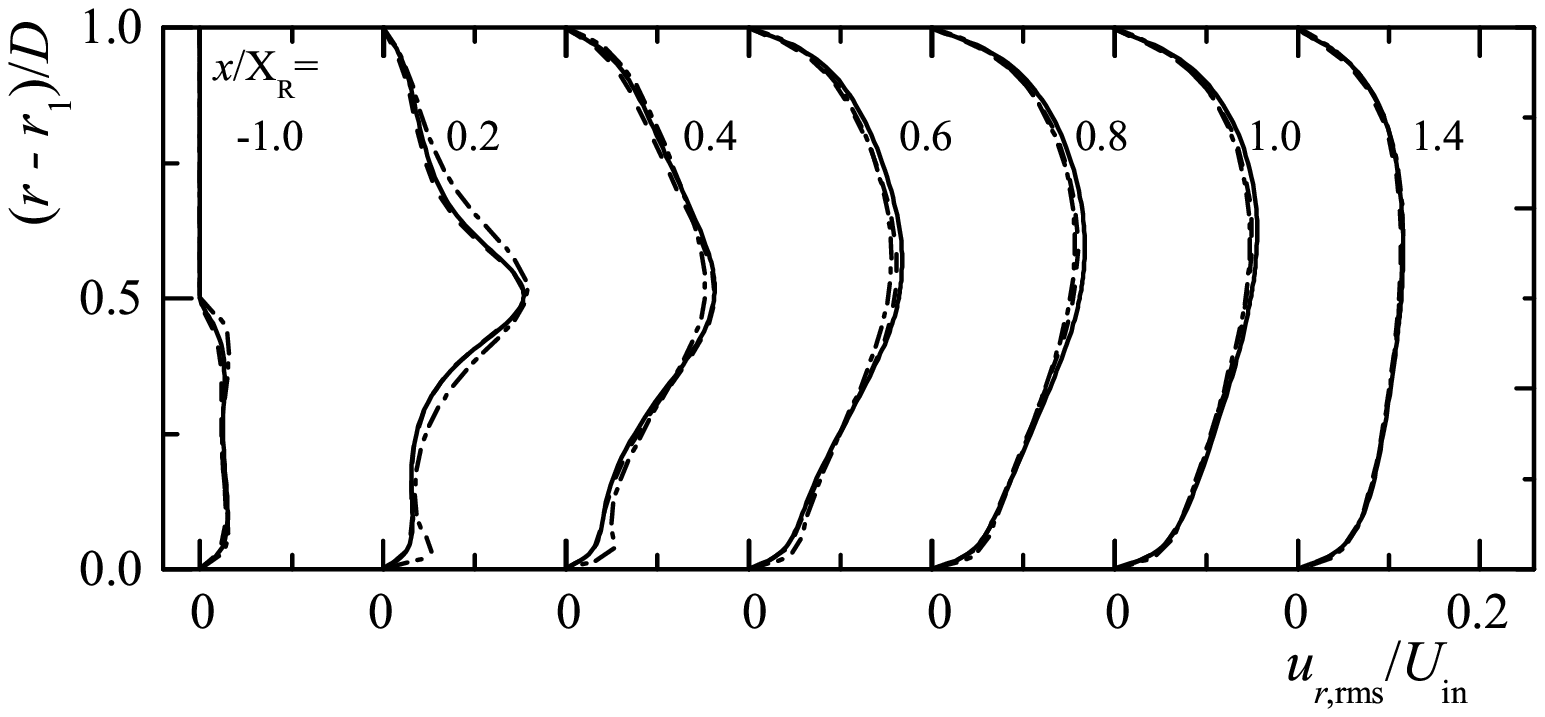} \\
(b) $u_{r,\mathrm{rms}}/U_\mathrm{in}$
\end{center}
\end{minipage}
\begin{minipage}{0.49\linewidth}
\begin{center}
\includegraphics[trim=0mm 7mm 0mm 0mm, clip, width=80mm]{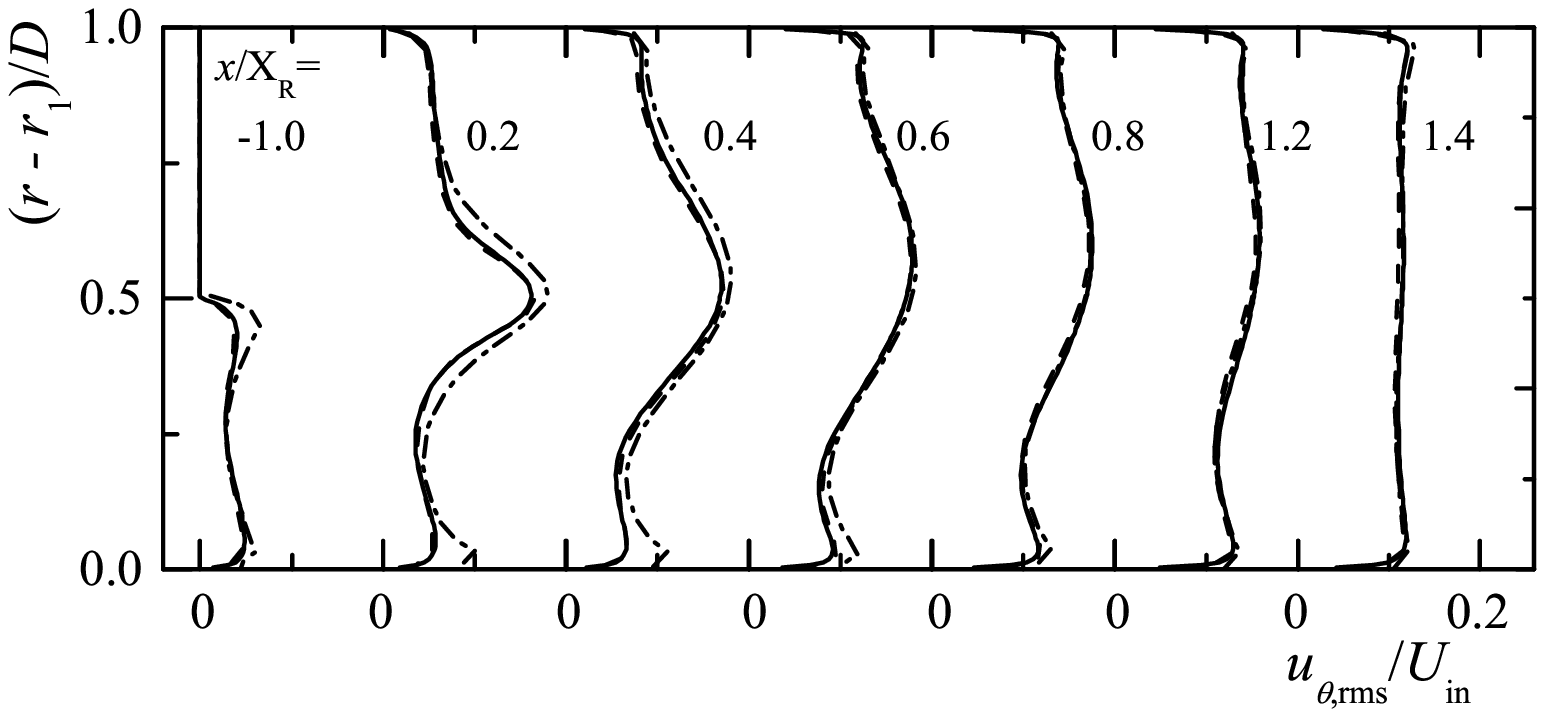} \\
(c) $u_{\theta,\mathrm{rms}}/U_\mathrm{in}$
\end{center}
\end{minipage}
\begin{minipage}{0.49\linewidth}
\begin{center}
\includegraphics[trim=0mm 7mm 0mm 0mm, clip, width=80mm]{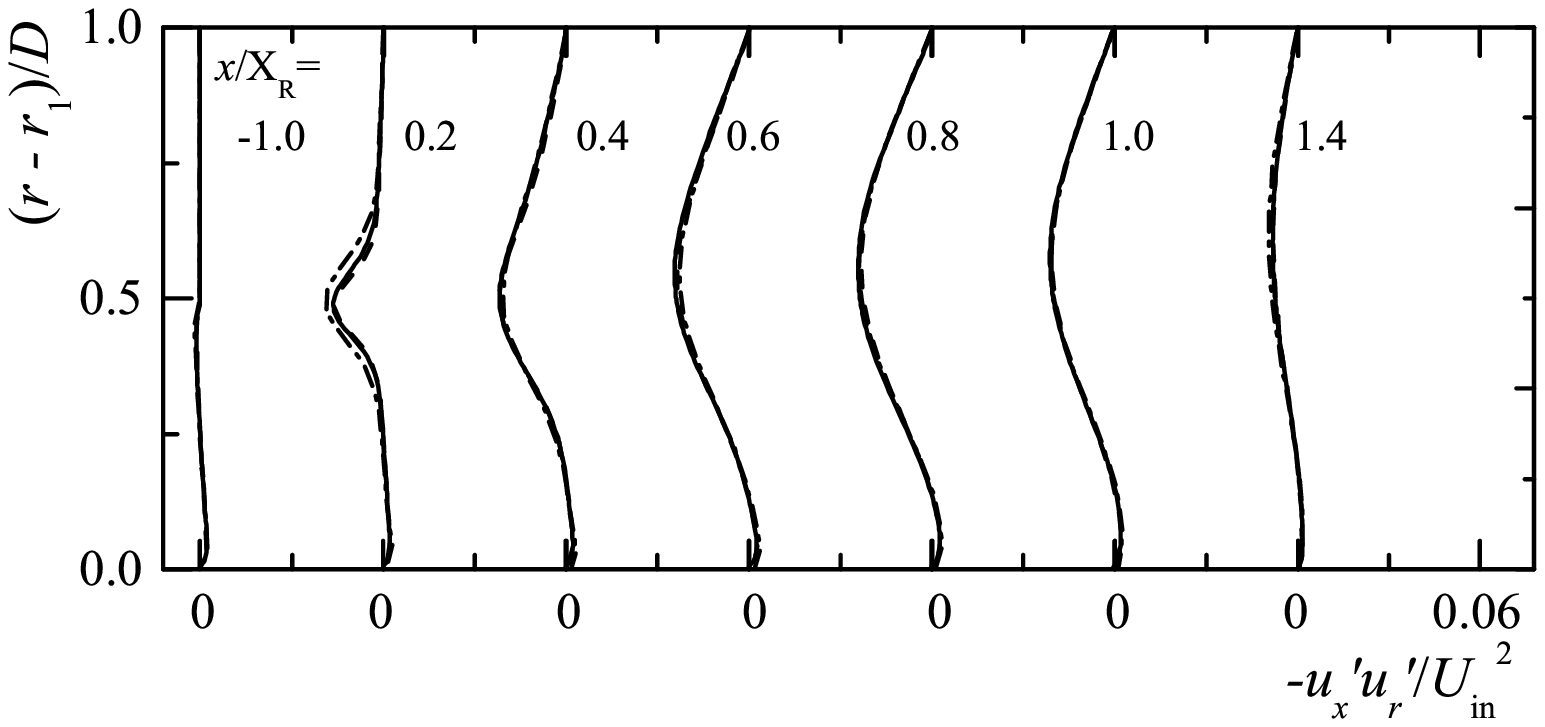} \\
(d) ${-u_x'}{u_r'}$
\end{center}
\end{minipage}
\caption{Turbulence intensity distributions of 
streamwise velocity, radial velocity, and circumferential velocity fluctuations, 
and Reynolds shear stress distribution for $\alpha=0.5$: 
- $\cdot$ -, grid1; - - -, grid2; ---, grid3.}
\label{rms_grid}
\end{figure}

\subsection{Turbulence characteristics}

Figure \ref{rms_ave} shows the turbulence kinetic energy $k$, 
the turbulence intensities of the streamwise velocity fluctuation 
$u_{x,\mathrm{rms}}$, radial velocity fluctuation $u_{r,\mathrm{rms}}$, 
and circumferential velocity fluctuation $u_{\theta,\mathrm{rms}}$. 
Each distribution was averaged over the $r-\theta$ cross-section. 
In the $k$ distribution shown in Fig. \ref{rms_ave} (a), 
for $\alpha=0.5$, 0.375, and 0.23, the turbulence becomes maximum 
near $x/D=3.2$, 3.49, and 4.2, respectively, 
due to longitudinal vortices similar to the rib structure existing in the flow field 
and small-scale vortices generated by the collapse of the vortex ring. 
Downstream, the turbulence decreases as the small-scale vortex group decay. 
In addition, the smaller $\alpha$, the more downstream the small vortices 
concentrated on the inner pipe side exist, 
so $k$ shows a high value even downstream from the position of maximum turbulence. 
This trend is the same in the turbulence intensity distribution of 
each velocity fluctuation.

Figurea \ref{rms} (a) to (c) show the turbulence intensity distribution of 
each velocity fluctuation. 
In all distributions, maximums exist around the shear layer at $(r-r_1)/D=0.5$ 
and near the inner pipe wall surface. 
We calculate the ratios of the maximum of $u_{r,\mathrm{rms}}$ and 
$u_{\theta,\mathrm{rms}}$ to $u_{x,\mathrm{rms}}$ around the separated shear layer 
at $x/X_\mathrm{R}=0.4$. 
The ratios are about 74\% and 76\% for $\alpha=0.5$, 
about 64\% and 68\% for $\alpha=0.375$, 
and about 55\% and 67\% for $\alpha=0.23$, respectively. 
Therefore, the flow has a strong three-dimensionality at this position 
because the longitudinal vortex existing around the vortex ring mixes 
the flow around the shear layer. 
Downstream, we can confirm that the turbulence of $0 \leq (r-r_1)/D \leq 0.5$ 
increases because the small-scale vortex group generated 
by the collapse of the vortex ring concentrates on the inner pipe side.
In the Reynolds stress distribution ${-u_x'}{u_r'}$ shown in Fig.\ref{rms} (d), 
regardless of $\alpha$, the magnitude of the extremum becomes maximum 
at $x/X_\mathrm{R}=0.6$ as in the $u_{x,\mathrm{rms}}$ distribution 
and attenuates downstream. The smaller $\alpha$, the smaller the extremum.

To show the validity of this calculation result, 
we compare the turbulence intensity distribution of the streamwise velocity fluctuation 
with the result of the three-dimensional backward step flow 
reported by \citet{Yanhua_et_al_2013}. 
In Fig. \ref{urms_datou}, $r_{\rm outer}=0$ is the outer pipe side wall surface, 
and $r_{\rm outer}=2h$ is the inner pipe side wall surface. 
Although $Re$ is different between this calculation and the experiment, 
the tendency that the maximum value due to the shear layer 
at $r_{\rm outer}=h$ gradually decreases toward the downstream is the same. 
In the present result, there is an extreme value near the wall surface 
due to the influence of the inner pipe wall surface. 
On the other hand, the maximum due to the wall surface cannot be seen 
near $r_{\rm outer}=2h$ in the experiment 
because the wall surface exists at about $r_{\rm outer}=80h$.

To confirm the dependency of the grid on the calculation result, 
we investigated the difference in the calculation result 
depending on the number of grid points for $\alpha=0.5$, 
where the grid resolution in the circumferential direction is the coarsest. 
Here, the distribution of turbulence averaged in the circumferential direction 
is shown. 
To evaluate the differences between the results obtained with grid1 and grid2 to grid3, 
we calculate the maximum difference mainly at the position 
where the distribution has a extreme value.

Figures \ref{rms_grid} (a) to (c) show the turbulence intensity distribution of 
each velocity fluctuation. 
Overall, there is a difference upstream of $x/X_\mathrm{R}=0.4$. 
In the distribution of $u_{x,\mathrm{rms}}$, 
when comparing the maximum values around $(r-r_1)/D=0.47$ at $x/X_\mathrm{R}=0.2$, 
the differences between the results of grid1 and grid2 to grid3 are 
about 6.7\% and 0.14\%, respectively. 
In the distribution of $u_{r,\mathrm{rms}}$, 
comparing the maximum values near $(r-r_1)/D=0.5$ at $x/X_\mathrm{R}=0.4$, 
the differences between grid1 and grid2 to grid3 are 
about 6.5\% and 0.19\%, respectively. 
In the distribution of $u_{\theta,\mathrm{rms}}$, 
comparing the maximum values near $(r-r_1)/D=0.5$ at $x/X_\mathrm{R}=0.2$, 
the differences between grid1 and grid2 to grid3 are 
about 10.7\% and 2.1\%, respectively. 
In the Reynolds stress distribution $-{u_x'}{u_r'}$ shown in Fig. \ref{rms_grid} (d), 
there is a difference at $x/X_\mathrm{R}=0.2$. 
Comparing the maximum value near $(r-r_1)/D=0.48$ at $x/X_\mathrm{R}=0.2$, 
the differences between the results of grid1 and grid2 to grid3 are 
about 17.3\% and 2.1\%, respectively. 
The difference between the results of grid2 and grid3 is small 
compared to the difference between the results of grid1 and grid2.

\section{Conclusions}

This study performed a large-eddy simulation of turbulent separated and reattached flow 
in an enlarged annular pipe. 
 The obtained findings are summarized as follows.
 
At all the pipe diameter ratios, the shear layer separated 
from the sudden expansion part becomes unstable 
and rolls up into a vortex, shedding a vortex ring periodically. 
A longitudinal vortex similar to the rib structure occurs 
around the vortex ring, making the flow three-dimensional. 
As a result, the vortex ring becomes unstable downstream 
and splits into small vortices. 
A tubular longitudinal vortex structure occurs downstream of the reattachment point 
near the wall surface on the inner pipe side.

A low-frequency fluctuation occurs at each pipe diameter ratio. 
This low-frequency fluctuation can promote the destabilization of the vortex ring 
and its collapse. 
The smaller the pipe diameter ratio is, 
the more downstream the influences of small-scale vortices 
and low-frequency fluctuation on the flow field appear.

The smaller the pipe diameter ratio, the slower the pressure recovery 
downstream from the reattachment point. 
In addition, the pressure recovery on the inner pipe side is delayed 
compared to the outer pipe side due to the tubular longitudinal vortex structure 
and small-scale vortices existing near the inner pipe wall.

Regardless of the pipe diameter ratio, turbulence is maximum 
upstream of the reattachment point due to the small-scale vortices 
generated by the collapse of the vortex ring. 
This maximum value decreases as the pipe diameter ratio decreases. 
In addition, the smaller the pipe diameter ratio, 
the higher the turbulence downstream from the reattachment point.

\section*{Acknowledgments}

The numerical results in this research were obtained 
using supercomputing resources at Cyberscience Center, Tohoku University. 
This research did not receive any specific grant from funding agencies 
in the public, commercial, or not-for-profit sectors. 
We would like to express our gratitude to Associate Professor 
Yosuke Suenaga of Iwate University for his support of our laboratory. 
The authors wish to acknowledge the time and effort of everyone involved 
in this study.

\section*{Additional Information}

\noindent
{\bf Declaration of Interests}: 
The authors report no conflict of interest.

\noindent
{\bf Author contributions}: 
H. Y. considered the content and policy of this research and constructed 
the calculation method and numerical codes. 
N. Y. performed the simulations. 
H. Y. and N. Y. contributed equally to analyzing data and reaching conclusions, 
and in writing the paper.

\noindent
{\bf Author ORCID}: H. Yanaoka https://orcid.org/0000-0002-4875-8174


\bibliographystyle{elsarticle-harv}
\bibliography{yamada_bibfile}

\end{document}